\documentclass[11pt,a4paper]{article}
\usepackage{graphicx}
\usepackage{amssymb}
\usepackage{amsmath}
\usepackage{amsfonts}
\usepackage{dsfont}
\usepackage{mathtools}
\usepackage{array}
\usepackage{subfig}
\usepackage{rotating}
\usepackage{bbold,amsfonts}
\usepackage{comment}
\usepackage[normalem]{ulem}

\usepackage[utf8]{inputenc}
\usepackage{bm}
\usepackage{xcolor}
\usepackage{float}
\usepackage{braket}
\usepackage{placeins}
\usepackage[height=8.8in,width=6.45in]{geometry}
\usepackage[font=small,labelfont=bf]{caption}
\usepackage[hidelinks]{hyperref}
\usepackage{booktabs,float,slashed}
\usepackage{standalone}
\usepackage{caption}
\usepackage{makecell}
\usepackage[maxbibnames=99,sorting=none,giveninits=true,backend=biber,style=numeric-comp,sortcites,doi=false,hyperref=true]{biblatex}
\renewbibmacro{in:}{}
\usepackage{hyperref}
\DeclareFieldFormat{doilink}{\iffieldundef{doi}{#1}{\href{https://doi.org/\thefield{doi}}{#1}}}
\DeclareFieldFormat[article,periodical]{volume}{\mkbibbold{#1}}
\DeclareFieldFormat[article,periodical]{journaltitle}{#1}
\DeclareFieldFormat[article,periodical]{pages}{#1}

\usepackage{xpatch}
\xpatchbibdriver{article}
  {\usebibmacro{journal+issuetitle}%
   \newunit
   \usebibmacro{byeditor+others}%
   \newunit
   \usebibmacro{note+pages}}
  {\printtext[doilink]{%
     \usebibmacro{journal+issuetitle}%
     \newunit
     \usebibmacro{byeditor+others}%
     \newunit
     \usebibmacro{note+pages}}}
  {}{}

\numberwithin{equation}{section}

\begin{document}
\begin{titlepage}
\vspace*{10mm}
\begin{center}
{\LARGE \bf 
Defect correlators in a $\mathcal{N}=2$ SCFT
at strong coupling
}

\vspace*{15mm}

{\Large Alessandro Pini and Paolo Vallarino}

\vspace*{8mm}
	
 Universit\`a di Torino, Dipartimento di Fisica\\
 and I.N.F.N. - sezione di Torino
			
			\vskip 0.3cm
						
   Via P. Giuria 1, I-10125 Torino, Italy

\vskip 0.8cm
	{\small
		E-mail:
		\texttt{apini,vallarin@to.infn.it}
	}
\vspace*{0.8cm}
\end{center}

\begin{abstract}
 We study the correlation function between one single-trace scalar operator and a circular Wilson loop in the $4d$ $\mathcal{N}=2$ superconformal field theory with gauge group $SU(N)$ and matter transforming in the symmetric and anti-symmetric representations. By exploiting supersymmetric localization, we resum the perturbative expansion of this correlator in the large-$N$ 't Hooft limit. Furthermore, using both analytical and numerical techniques, we provide a prediction for the leading term of its strong coupling expansion and we compare this prediction to numerical Padé resummations of the perturbative series.

\end{abstract}
\end{titlepage}
\setcounter{tocdepth}{2}
\tableofcontents

\section{Introduction}
\label{sec:intro}
Four dimensional gauge theories with high amount of supersymmetry are a very fruitful field of study since they provide a formidable source for exact results in quantum field theories and their analysis can shed light on our understanding of the strong coupling regime. Along the years, the study of these theories has been mostly carried out using integrability, supersymmetric localization and the AdS/CFT correspondence. In particular, relevant progress has been made in characterizing  the planar limit of the maximally supersymmetric quantum field theory in four dimensions, i.e. $\mathcal{N}=4$ Super Yang-Mills (SYM).

Although in general more difficult, the study of gauge theories with non maximal supersymmetry is interesting as well. During the last years a significant amount of progress has been achieved in the context of $4d$ $\mathcal{N}=2$ gauge theories by exploiting supersymmetric localization. This computational technique permits to replace a $\mathcal{N}=2$ SYM theory on flat space-time with an interacting matrix model on a 4-sphere $S^4$. This way the computation of the vacuum expectation value of an observable in the $\mathcal{N}=2$ SYM theory which, in principle, should be performed by computing an infinite dimensional path integral, can be carried out via the evaluation of a finite dimensional integration over the elements of a matrix. 
When the $\mathcal{N}=2$ theory is conformal, this method led to several results such as the computation of correlation functions among chiral and anti-chiral scalar operators \cite{Gerchkovitz:2016gxx,Baggio:2016skg,Rodriguez-Gomez:2016ijh,Rodriguez-Gomez:2016cem,Pini:2017ouj,Fiol:2021icm,Billo:2022gmq,Billo:2022fnb,Billo:2022lrv,Galvagno:2020cgq,Baggio:2015vxa,Beccaria:2020hgy,Billo:2022xas,Beccaria:2021hvt,Bobev:2022grf,Beccaria:2022ypy}, the vacuum expectation value of Wilson loops \cite{Billo:2019fbi,Beccaria:2021ksw,Beccaria:2021vuc,Beccaria:2021ism,Passerini:2011fe,Zarembo:2020tpf} and the free energy of the theory \cite{Fiol:2021jsc,Fiol:2020bhf}.

A particular $\mathcal{N}=2$ theory that turned out to be particularly suitable to be analysed by the application of localization techniques is the so called \textbf{E}-theory \cite{Beccaria:2020hgy,Beccaria:2021hvt}. This is a superconformal gauge theory with gauge group $SU(N)$ and matter transforming in the symmetric and anti-symmetric representations of the gauge group. Remarkably, although it is an $\mathcal{N}=2$ theory, in the planar limit it shares many properties with $\mathcal{N}=4$ SYM. As a matter of fact the difference among the two central charges $a$ and $c$ is zero at the leading order of the large-$N$ expansion. Furthermore many observables of the \textbf{E}-theory are planar equivalent to $\mathcal{N}=4$ SYM. Examples of such observables are provided by the expectation value of a circular Wilson loop, the free energy and the correlators of single-trace chiral scalar operators with even conformal dimensions \cite{Beccaria:2020hgy,Beccaria:2021vuc,Beccaria:2021hvt}. Finally the \textbf{E}-theory admits a gravity dual given by a suitable $\mathbb{Z}_2$ orbifold/orientifold projection of the  $\mathcal{N}=4$ SYM gravity dual \cite{Ennes:2000fu}. 

In this article we aim to further study the \textbf{E}-theory and move a new step towards a more complete understanding of its planar limit and strong coupling regime. As mentioned above, crucial to this study is the use of the corresponding matrix model that differs from the matrix model of the $\mathcal{N}=4$ theory by a non-trivial interaction action and a term containing non-perturbative instanton contributions. Nevertheless a lot of simplifications occurs in the planar limit, where the non-perturbative contributions become negligible and the partition function of the matrix model is given by
\begin{align}
    \mathcal{Z} = \textrm{det}^{-\frac{1}{2}}(1-\mathsf{X})\, \ ,
\end{align}
where $\mathsf{X}$ is a semi-infinite matrix whose elements are a convolution among Bessel functions of the first kind. All the dependence on the 't Hooft coupling $\lambda$ only enters via the Bessel functions. Due to this, the $\mathsf{X}$-matrix turns out to be a very efficient computational tool. The weak coupling regime can be analysed by expanding the Bessel functions in power series of the 't Hooft coupling. This way it is computationally easy to generate very long perturbative expansions, that can then be used for the numerical evaluation of the expectation value of many observables (see for instance \cite{Beccaria:2020hgy}). On the other hand, by exploiting the properties of the $\mathsf{X}$-matrix for large values of $\lambda$, we can extract information on the strong coupling regime of the theory (see \cite{Beccaria:2021hvt}).

Here we develop a similar analysis for a defect correlation function between one single-trace chiral scalar operator and a circular Wilson loop in the fundamental representation. The planar limit of this correlation function for the maximally supersymmetric theory has been performed years ago (see for example \cite{Beccaria:2020hgy}). 
However, in a $\mathcal{N}=2$ context, this observable has received less consideration. In the matrix model literature the only available results have been obtained at the perturbative level for some specific theories such as SQCD \cite{Billo:2018oog}, quiver gauge theories \cite{Galvagno:2021bbj,Preti:2022inu} or other (in general non conformal) $\mathcal{N}=2$ theories \cite{Sysoeva:2017fhr}. To the best of our knowledge, for $\mathcal{N}=2$ superconformal theories, no exact results in the planar limit are known. In this paper we fill this gap for the \textbf{E}-theory. By exploiting the properties of its matrix model we analytically derive an expression of this 1-point function valid in the large-$N$ limit for any value of the 't Hooft coupling. Then, based on this expression and by exploiting the properties of the above mentioned $\mathsf{X}$ matrix, we provide a prediction for the leading term of its strong coupling expansion.

This paper is organised as follows: in Section \ref{sec:wilsonloop} we review the main features of the \textbf{E}-theory and of the correlation function under our analysis in the general framework of a $4d$ $\mathcal{N}=2$ SCFT. In Section \ref{sec:loc} we review the most relevant aspects of the matrix model for the $\mathcal{N}=4$ SYM and then for the \textbf{E}-theory, also recalling how quantum field theory operators are represented in the matrix model. Then, as warm up, in Section \ref{sec:example} we consider the simplest correlation function of interest for us, i.e. the one involving a chiral operator of conformal dimension three. This case is particularly suitable since it permits to outline the procedure for the resummation of the perturbative series without the need to deal with the technicalities of the Gram-Schimdt orthogonalization. Exploiting the properties of the modified Bessel functions at strong coupling
as well as the numerical Padé resummation of its perturbative series we provide a conjecture for the leading term of the strong coupling expansion of this correlator. In Section \ref{sec:general} we perform the same analysis for the most general correlator and provide an explicit expression for the corresponding strong coupling expansion. Then we validate it through some numeral checks.
Finally we draw our conclusions in Section \ref{sec:conclusions}.

\section{Correlator among one chiral operator and a Wilson loop}
\label{sec:wilsonloop}
We consider the $4d$ $\mathcal{N}=2$ superconformal field theory with gauge group $SU(N)$ on $\mathbb{R}^{4}$ whose matter content consists of one hypermultiplet transforming in the symmetric representation of the gauge group and one hypermultiplet transforming in the anti-symmetric one. This theory has a vanishing 1-loop $\beta$ function coefficient and, therefore, is conformal.  We denote by $g$ the corresponding coupling constant and by $\lambda \equiv Ng^2$ the 't Hooft coupling.

We consider the complex scalar $\varphi(x)$ inside the $\mathcal{N}=2$ vector multiplet and we construct the set of local, scalar, gauge invariant operators 
\begin{align}
O_{\textbf{n}}(x) \equiv \textrm{tr} \varphi^{n_1}(x) \ \textrm{tr} \varphi^{n_2}(x) \ \cdots \ \textrm{tr} \varphi^{n_\ell} (x) \, , 
\label{eq:O_operator}
\end{align}
where $\textbf{n} =\{n_1,n_2,\dots, n_{\ell}\}$. The operators \eqref{eq:O_operator} are chiral conformal primary operators and their conformal dimension $\Delta_{\textbf{n}}$ is protected and only depends on the $U(1)_r$ R-charge. In our conventions it is given by
\begin{align}
    \Delta_{\textbf{n}} = \sum_{i=1}^{\ell} n_i \, .
\end{align}

The other gauge invariant operator that we consider is the half-BPS  Wilson loop in the fundamental representation along a circle $C$ of radius $R$ \cite{Maldacena:1998im,Berenstein:1998ij,Semenoff:2001xp}. We denote such operator by $W_C$ and its explicit expression is given by
\begin{align}
    W_C \equiv \frac{1}{N} \ \textrm{tr} \ \mathcal{P} \ \textrm{exp} \left\{ g \oint_{C} d\tau \, \left[i A_{\mu}(x)\dot{x}^{\mu}(\tau) + \ \frac{R}{\sqrt{2}}\left(\varphi(x)+\bar{\varphi}(x)\right)\right] \right\}\, ,
\label{Wilsonloop}    
\end{align}
where $A_{\mu}(x)$ is the gauge field and $\mathcal{P}$ denotes the path-ordering. Without any loss of generality, we choose to place the circle $C$ inside the $\mathbb{R}^2 \subset \mathbb{R}^4$ parameterized by $(x_1,x_2)$. This way the explicit expression of the function $x^{\mu}(\tau)$ appearing in \eqref{Wilsonloop} is
\begin{align}
    x^{\mu}(\tau) = R \ (\cos \tau, \sin \tau, 0, 0),  \ \ \ \ \ 0 \leq \tau < 2\pi \, .
\end{align}

In this work we consider the correlator between the circular Wilson loop in \eqref{Wilsonloop} with $R=1$ and one chiral scalar operator \eqref{eq:O_operator}. This quantity is strongly constrained by conformal invariance and reads \cite{Billo:2016cpy,Billo:2018oog}
\begin{align}
\langle W_C \ O_\textbf{n}(x) \rangle =\frac{w_\textbf{n}(g,N)}{(2\pi||x||_C)^{\Delta_{\textbf{n}}}}\, ,
\label{1point}
\end{align}
where $||x||_C$ denotes the chord distance\footnote{See equation (2.10) of \cite{Billo:2018oog} for the definition of this distance.} between the circular Wilson loop and the insertion point of the chiral operator $O_\textbf{n}(x)$. Therefore, the only quantity that must be determined is the function $w_\textbf{n}(g,N)$ appearing at the numerator of \eqref{1point}, which depends non-trivially on the coupling $g$ and the rank of the gauge group. 

Henceforth we focus on the particular case of a single-trace chiral operator $O_n(x)$. Our aim is to determine an exact expression for $w_n(g,N)$ valid, in the large-$N$ limit, for any value of the 't Hooft coupling $\lambda$ and provide a conjecture for the leading term of its strong coupling expansion. In the next section we begin by reviewing the matrix model used for the evaluation of the function $w_n(g,N)$.

\section{Localization}
\label{sec:loc}

Supersymmetric localization allows to map the computation of correlators in $\mathbb{R}^4$ to an interacting matrix model defined on a 4-sphere $S^4$ \cite{Pestun:2007rz} (for a review see \cite{Pestun:2016jze}). This way the calculation of the coefficients $w_n(g,N)$ is reduced to the evaluation of a finite-dimensional matrix integral.

\subsection{Matrix model}

In the ``full Lie algebra approach" \cite{Billo:2017glv}, the partition function $\mathcal{Z}$ can be written as an integral over all components of a $N \times N$ Hermitian traceless matrix $a$, such that
\begin{align}
a=a^b\,T_b\, , \ \ \ \ \ \ \ \ \ \ b=1,\dots,N^2-1 ~,
\end{align}
where $T_b$ are the $\mathfrak{su}(N)$ generators in the fundamental representation, so that
\begin{align}
\textrm{tr} \,T_b\,T_c = \frac{1}{2}\,\delta_{b,c}\, , \ \ \ \ \ \ b,c=1,\dots,N^2-1 ~.
\end{align}
Hence the partition function takes the form
\begin{align}
    \mathcal{Z} = \int \!da \ \textrm{e}^{-\textrm{tr} \, a^2}\, |Z_{1-loop}\, Z_{inst}|^2 ~,
    \label{eq:Z}
\end{align}
where the integration measure is defined as 
\begin{align}
    da = \prod_{b=1}^{N^2-1}\frac{da^b}{\sqrt{2\pi}}~.
\end{align}
In the planar limit, the instanton contributions are exponentially suppressed and so we can set $Z_{inst}=1$. On the other hand, $Z_{1-loop}$ is the contribution due to the 1-loop determinants of the fluctuations around the localization locus which can be written in terms of an interaction action as
\begin{align}
|Z_{1-loop}|^2=e^{-S_{int}}~.
\end{align}
Therefore the vacuum expectation value of any function $f(a)$ becomes
\begin{align}
\langle f(a) \rangle = \frac{\int \!da \ \textrm{e}^{-\textrm{tr} \, a^2-S_{int}}\, f(a)}{\int \!da \ \textrm{e}^{-\textrm{tr} \, a^2-S_{int}}}\, =\, \frac{\langle e^{-S_{int}}\, f(a) \rangle_0}{\langle e^{-S_{int}} \rangle_0}~,
\end{align}
where $\langle\, \rangle_0$ stands for the expectation value in the free matrix model.

\subsection{The operators in the matrix model}

We now recollect how the gauge theory operators introduced in Section \ref{sec:wilsonloop} are represented in the matrix model. 

First of all, in the matrix model it is natural to introduce the multi-trace operators
\begin{align}
\label{eq:A_operators}
A_{\textbf{n}}\equiv\textrm{tr}\,a^{n_1}\textrm{tr}\,a^{n_2}\cdots\textrm{tr}\,a^{n_{\ell}}\qquad\mbox{with}~\textbf{n} =\{n_1,n_2,\dots, n_{\ell}\}\, .
\end{align}
However, these operators do not directly correspond to the operators $O_\textbf{n}(x)$ defined in \eqref{eq:O_operator}. In order to obtain the correct representation of these latter in the matrix model, it is necessary to perform a Gram-Schmidt orthogonalization procedure, that allows to subtract from \eqref{eq:A_operators} all its contractions with operators of lower dimensions \cite{Gerchkovitz:2016gxx}. Thus, if we conveniently introduce the vevless operators
\begin{align}
\hat{A}_{\textbf{n}} \equiv A_{\textbf{n}}-\langle A_{\textbf{n}} \rangle,
\end{align}
the gauge invariant operators in the matrix model are obtained as follows
\begin{align}
\label{eq:changebasis}
O_\textbf{n}=\sum_{\textbf{m} \leq \textbf{n}}\,M_{\textbf{n},\textbf{m}}\,\hat{A}_\textbf{m}~,
\end{align}
where the mixing  Gram-Schimdt coefficients $M_{\textbf{n},\textbf{m}}$, which in general have a non-trivial dependence on $\lambda$, are constructed by requiring the orhogonality among $O_\textbf{n}$ and all  the operators of lower dimension. \\
It is worth noting that, since the matrices $a$ take values in the $\mathfrak{su}(N)$ Lie algebra, it follows that 
\begin{align}
\label{Aodd}
\langle A_{\textbf{n}} \rangle = 0  \qquad\mbox{if}~\Delta_{\textbf{n}}~\mbox{is odd} \, ,
\end{align}
which implies that there is no mixing among one odd and one even single-trace operator. 

For simplicity, here we illustrate how this orthogonalization procedure works in the case of the first single-trace scalar operators.\footnote{For the aim of this paper, we are just interested in correlators that involve 2-point functions of chiral scalar operators, for which in \cite{Baggio:2016skg,Billo:2022xas} it was shown that in the large-$N$ limit it is sufficient to carry out the Gram-Schmidt orthogonalization on the single-trace operators only.} According to \eqref{eq:changebasis}, we get\footnote{Note that from \eqref{Aodd} follows that $\hat{A}_{\textbf{n}}=A_{\textbf{n}}$ if $\Delta_\textbf{n}$ is odd.}
\begin{align}
& O_2=\hat{A}_2 \, ,\quad \ \, \qquad \qquad \qquad \qquad\ O_3=A_3 \, ,  \notag \\
& O_4=\hat{A}_4 -\frac{\langle \hat{A}_4\, \hat{A}_2 \rangle}{\langle \hat{A}_2\, \hat{A}_2 \rangle}\hat{A}_2\, , \qquad \qquad O_5=A_5 -\frac{\langle A_5\, A_3 \rangle}{\langle A_3\, A_3 \rangle}\,A_3 \, . 
\label{Mnm}
\end{align}
In general it is possible to derive a closed expression valid for the generic $M_{n,m}$ coefficient as shown in \cite{Billo:2022xas}.

On the other hand, in the matrix model the $1/2$ BPS circular Wilson loop of unitary radius in the fundamental representation is embodied by the insertion of the following operator \cite{Pestun:2007rz} 
\begin{equation}
\label{WC} 
W_C = \frac{1}{N}\textrm{tr}\, \textrm{exp} \biggl( \frac{g}{\sqrt{2}}\,a \biggr) = \frac{1}{N}\sum_{k=0}^{\infty}\, \frac{g^k}{2^{\frac{k}{2}}\, k!}\, \textrm{tr}a^{k}~.
\end{equation}

\subsection{The \texorpdfstring{$\mathcal{N}=4$}{} SYM theory}

Here we briefly review the main results obtained in the $\mathcal{N}=4$ SYM theory in the 't Hooft limit. In this case the matrix model is non-interacting, i.e. it is purely Gaussian. 

As found in \cite{Rodriguez-Gomez:2016cem}, in the large-$N$ limit, it is possible to find a closed form for the mixing coefficients defined in \eqref{eq:changebasis}, valid for single-trace operators with conformal dimension $n$ \footnote{The symbol $\simeq$ means that the equality holds at the planar level of the large-$N$ expansion. We will use this notation also in the following.} 
\begin{align}
\label{N=4mixing}
O_{n}^{(0)}\simeq n\sum_{k=0}^{\lfloor\frac{n-1}{2}\rfloor}(-1)^k\biggl(\frac{N}{2} \biggr)^k\frac{(n-k-1)!}{k!(n-2k)!}\hat{A}_{n-2k}\, ,
\end{align}
where we denoted with the label $(0)$ the normal-ordered operators in the free matrix model. 

From \eqref{N=4mixing}, it is quite straightforward to derive the 2-point function between one chiral and one anti-chiral scalar operator, that reads \cite{Beccaria:2020hgy}
\begin{align}
\label{Gn}
    \langle O_n^{(0)}\, \overline{O}_n^{(0)} \rangle_{0} \simeq n\left(\frac{N}{2}\right)^{n} \equiv \mathcal{G}_n\, .
\end{align}
Finally, it is useful to recall the result found for the correlator among a circular Wilson loop in the fundamental representation and one chiral primary operator. Exploiting the \eqref{WC} and the \eqref{N=4mixing}, in the planar limit the expression for this correlator becomes
\begin{align}
\label{OWfree}
\langle O_n^{(0)} \, W_C \rangle_{0} \simeq \frac{n}{N}\sum_{k=0}^{\infty}\,\frac{1}{k!} \biggl(\frac{\lambda}{2N}\biggr)^{\frac{k}{2}}\sum_{i=0}^{\lfloor \frac{n-1}{2}\rfloor}(-1)^i\biggl(\frac{N}{2} \biggr)^i\frac{(n-i-1)!}{i!(n-2i)!}\langle\hat{A}_k\,\hat{A}_{n-2i}\rangle_0\, .
\end{align}
The expression for the 2-point function of non normal-ordered operators in the free matrix model $\langle \hat{A}_p\,\hat{A}_{\ell}\rangle_0$  was derived in \cite{Beccaria:2020hgy}. Inserting it in \eqref{OWfree} we get
\begin{align}
\label{N=4}
    \langle O_n^{(0)} \, W_C \rangle_{0} \simeq \frac{\sqrt{n\,\mathcal{G}_n}}{N}I_{n}(\sqrt{\lambda}) \, ,
\end{align}
where $I_n(\sqrt{\lambda})$ is a modified Bessel function of the first kind. The expression \eqref{N=4} agrees with the well-known result obtained in \cite{Semenoff:2001xp} by resumming rainbow diagrams in the planar limit. 

\subsection{The interacting \textbf{E}-theory}

Given its matter content and exploiting supersymmetric localization it can be shown that the interaction action for the \textbf{E}-theory reads \cite{Beccaria:2020hgy}
\begin{align}
\label{Sint}
S_{\mathrm{int}} = 
2\sum_{m=1}^{\infty}\sum_{k=1}^{m-1}(-1)^{m}\Big(\frac{\lambda}{8\pi^2N}\Big)^{m+1}\,\binom{2m+2}{2k+1}\,
\frac{\zeta_{2m+1}}{m+1}\,\textrm{tr} \, a^{2k+1}\, \textrm{tr}\, a^{2m-2k+1} \, , 
\end{align}
where $\zeta_{2m+1}$ is the Riemann $\zeta$-value $\zeta(2m+1)$.  We observe that the expression \eqref{Sint} implies that the first non-trivial perturbative contributions for the \textbf{E}-theory is proportional to $\zeta_5$, while in others $\mathcal{N}=2$ SCFTs (e.g. in SQCD) the first term of the corresponding interaction action is proportional to $\zeta_3$. The diagrammatic counterparts of this difference is realized in the fact that the first relevant Feynman diagrams for the \textbf{E}-theory are 1-loop higher than other $\mathcal{N}=2$ SCFTs.\\
Furthermore we notice that the expression \eqref{Sint} is quadratic in traces of $a$ and only odd powers appear. These properties have been exploited in \cite{Beccaria:2020hgy} to show that, in the large-$N$ limit, the partition function $\mathcal{Z}$ for the \textbf{E}-theory matrix model drastically simplifies. To make this explicit, let us introduce the new operators
\begin{align}
\omega_k = \frac{O_{2k+1}^{(0)}}{\sqrt{\mathcal{G}_{2k+1}}} \, ,
\end{align}
whose two point function in the free theory at large-$N$ is canonically normalized
\begin{align}
\langle \omega_k\,\omega_{\ell} \rangle_{0} = \delta_{k,\ell} \, . 
\end{align}
If we introduce the infinite column vector
\begin{align}
    \boldsymbol{\omega}^{T} \equiv \ (\omega_{1},\omega_{2},\dots )\, , 
\end{align}
the interaction action \eqref{Sint} can be rewritten as
\begin{align}
S_{\mathrm{int}}(a) = -\frac{1}{2} \ \boldsymbol{\omega}^T \ \mathsf{X} \ \boldsymbol{\omega}\, ,
\end{align}
where $\mathsf{X}$ is a semi-infinite symmetric matrix whose entries are given by a convolution of Bessel functions of the first kind
\begin{align}
\label{Xmatrix}
\mathsf{X}_{k,\ell} = -8(-1)^{k+\ell}\sqrt{(2k+1)(2\ell+1)}\int_{0}^{\infty}\!\frac{dt}{t}\frac{e^{t}}{(e^t-1)^2}\ J_{2k+1}\Big(\frac{t\sqrt{\lambda}}{2\pi}\Big)J_{2\ell+1}\Big(\frac{t\sqrt{\lambda}}{2\pi}\Big)~,
\end{align}
for $k,\ell \geq 1$. Importantly all dependence on the 't Hooft coupling $\lambda$ has been summed up in terms of the Bessel functions. If we Taylor expand the Bessel functions and we analytically perform the integration over $t$ we recover the perturbative expression \eqref{Sint}. Moreover the expression \eqref{Xmatrix} can be used to get information at strong coupling. This can be achieved using the inverse Mellin transform of two Bessel functions and exploiting its asymptotic expansion for $\lambda >>1$. We finally observe that the partition function of the matrix model $\mathcal{Z}$ can be written in terms of the $\mathsf{X}$-matrix as
\begin{align}
    \mathcal{Z} = \int D\boldsymbol{\omega} \  \textrm{e}^{-\frac{1}{2}\,\boldsymbol{\omega}^T(\mathbb{1}-\mathsf{X})\,\boldsymbol{\omega}} =  \textrm{Det}^{-\frac{1}{2}}(\mathbb{1}-\mathsf{X})\, ,
\end{align}
 where
 \begin{align}
     D \boldsymbol{\omega} = \prod_{k=1}^{\infty}\frac{d\omega_k}{\sqrt{2\pi}} \, .
 \end{align}

Since we need them in the following sections, we review the properties of two-point correlator $\langle A_k A_{\ell} \rangle$ in the planar limit of the interacting $\textbf{E}$-theory. Given the particular expression of the interaction action \eqref{Sint}, it follows that
\begin{align}
    \langle A_{2k}\,A_{2\ell} \rangle \simeq \langle A_{2k}\,A_{2\ell} \rangle_0\, , \ \ \ \ \ \ \langle A_{2k}\,A_{2\ell+1} \rangle \simeq 0\, , 
\end{align}
while the 2-point correlator among two odd-dimensional traces has a more involved expression and  reads \cite{Beccaria:2020hgy}
\begin{align}
\label{eq:AANgrande}
\langle A_{2n+1}\,A_{2m+1} \rangle \simeq \biggl( \frac{N}{2} \biggr)^{n+m+1}\,\sum_{i=0}^{n-1}\sum_{j=0}^{m-1}\, c_{n,i}\, c_{m,j}\,\mathsf{D}_{n-i,m-j} ~, 
\end{align}
where the numerical coefficients $c_{n.i}$ are given by
\begin{align}
c_{n,i}=\binom{2n+1}{i}\sqrt{2n+1-2i} ~.
\end{align}
We stress that all the dependence on the 't Hooft coupling in \eqref{eq:AANgrande} enters only through $\mathsf{D}_{i,j}$, that can be completely expressed in terms of the $\textsf{X}$-matrix as follows
\begin{equation}
\mathsf{D}_{i,j}=\left(\frac{1}{\mathbb{1}-\mathsf{X}}\right)_{i,j}=\delta_{i,j}+\mathsf{X}_{i,j}+(\mathsf{X}^2)_{i,j}+(\mathsf{X}^3)_{i,j}+\dots \, \ .
\label{D}
\end{equation}
This permits to study the properties of the two point correlators \eqref{eq:AANgrande} at strong coupling and, at the same time, to generate very long perturbative series. This last procedure can be performed using the following identity among Bessel functions 
\begin{align}
    G(t_1,t_2) = 8\sum_{k=1}^{\infty}(2k+1)J_{2k+1}(t_1)J_{2k+1}(t_2) = -\frac{4t_1t_2}{t_1^2-t_2^2}\left(t_1J_{1}(t_1)J_2(t_2)-t_2J_2(t_1)J_1(t_2)\right)\, ,
\end{align}
this way the first powers of the $\textsf{X}$-matrix entering in \eqref{D} read
\begin{align}
\label{X2}
& (\textsf{X}^2)_{i,j} = 8(-1)^{i+j}\sqrt{(2i+1)(2j+1)}\int \mathcal{D}t \ \mathcal{D}t' \ J_{2i+1}\left(zt\right) G\left(zt, zt' \right) J_{2j+1}\left(zt'\right)\, \ , \\
& (\textsf{X}^3)_{i,j} = -8(-1)^{i+j}\sqrt{(2i+1)(2j+1)}\int \mathcal{D}t \ \mathcal{D}t' \ \mathcal{D}t^{''} J_{2i+1}(zt)\,G(zt,zt')\,G(zt',zt^{''})\, J_{2j+1}(zt^{''}) \, , \nonumber
\end{align}
where we used the notation
\begin{align}
 \mathcal{D}t = \frac{dt}{t}\frac{\textrm{e}^{t}}{(\textrm{e}^{t}-1)^2}, \ \ \ \   z = \frac{\sqrt{\lambda}}{2\pi}\, . 
\end{align}
For example it is easy to generate the perturbative expansions for the compenents $i=j=1$
\begin{align}
    & \textsf{X}_{1,1} = -\frac{5\,\zeta_5\,\lambda^3}{256\,\pi^6} + \frac{105\,\zeta_7\,\lambda^4}{4096\,\pi^8} - \frac{1701\,\zeta_9\,\lambda^5}{65536\,\pi^{10}} + \frac{12705\,\zeta_{11}\,\lambda^6}{524288\,\pi^{12}} - \frac{184041\,\zeta_{13}\,\lambda^7}{8388608\,\pi^{14}} + O (\lambda^8)\, , \label{X11pert} \\
    & (\textsf{X}^2)_{1,1} = \frac{25\,\zeta_5^2\,\lambda^6}{65536\,\pi^{12}} - \frac{525\,\zeta_5\,\zeta_7\,\lambda^7}{524288\,\pi^{14}} + \frac{\lambda^8}{\pi^{16}} \biggl( \frac{44835\,\zeta_7^2}{67108864} + \frac{8505\,\zeta_5\,\zeta_9}{8388608} \biggr) + O (\lambda^9)\, , \\
    & (\textsf{X}^3)_{1,1} = -\frac{125\,\zeta_5^3\,\lambda^9}{16777216\,\pi^{18}} + \frac{7875\,\zeta_5^2\,\zeta_7\,\lambda^{10}}{268435456\,\pi^{20}} - \frac{\lambda^{11}}{\pi^{22}} \biggl( \frac{334425\,\zeta_5\,\zeta_7^2}{8589934592} + \frac{127575\,\zeta_5^2\,\zeta_9}{4294967296} \biggr) + O (\lambda^{12})\, \label{X311pert}.
\end{align}

 We finally define another quantity that will be useful in the following, which was firstly introduced in \cite{Billo:2022fnb} in the study of 3-point functions of chiral primary operators, namely the coefficient $\mathsf{d_n}$  
\begin{equation}
    \mathsf{d}_n=\sum_{n^\prime}\sqrt{n^\prime}\,\mathsf{D}_{n,n^\prime} ~,
    \label{dk}
\end{equation}
whose strong-coupling limit reads \cite{Billo:2022fnb}
\begin{align}
\label{dstrong}
\mathsf{d}_{n} \underset{\lambda\rightarrow\infty}{\sim} \frac{2\pi}{\sqrt{\lambda}}\sqrt{(2n+1)}(n^2+n)\,+\,O\biggl( \frac{1}{\lambda}\biggr) ~.
\end{align}

\subsection{The defect correlator}
We now consider the defect correlator  $\langle O_k\,W_C \rangle$ in the $\textbf{E}$-theory. When we turn off the interaction, we have to recover the $\mathcal{N}=4$ result \eqref{N=4}. For this reason we find convenient to rewrite the planar limit of this correlator as 
\begin{align}
\label{target}
    \langle O_k\, W_C \rangle \simeq \langle O_k\,W_C \rangle_0\,\biggl(1+\Delta w_k(\lambda)\biggr)\, , 
\end{align}
where $\Delta w_k(\lambda)$ is a non-trivial function of the 't Hooft coupling that must satisfy 
\begin{align}
    \lim_{\lambda \to 0 }\Delta w_k(\lambda) = 0 \, .
\end{align}
In order to compute \eqref{target} we first expand the operator $O_k$ on the basis of the single-traces $\{ A_\ell \}$ \eqref{eq:A_operators} as
\begin{align}
 O_k = \sum_{\ell=1}^{k}M_{k,\ell}\,A_{\ell} \, .  
\end{align}
We now treat separately the cases in which $k$ is even or odd. In the first case, since 
\begin{align}
 \langle A_{2p}\,A_{2\ell} \rangle \simeq \langle A_{2p}\,A_{2\ell} \rangle_0\, ,   
\end{align}
the Gram-Schmidt coefficients $M_{2p,2\ell}$ coincide with those of the free theory. Therefore we conclude that
\begin{align}
    \Delta w_{2p}(\lambda) = 0\, .
\end{align}
This implies that the insertion of a chiral operator with even conformal dimension $k=2p$ is planar equivalent to $\mathcal{N}=4$ SYM. On the other hand, for $k=2p+1$ the Gram Schmidt coefficients $M_{2p+1,2\ell+1}$ depend on the 't Hoof coupling in a non-trivial way. Therefore, this time, we are left with the evaluation of
\begin{align}
\label{START}
\langle O_{2p+1}\,W_C \rangle \simeq \sum_{\ell=1}^{p}M_{2p+1,2\ell+1}(\lambda) \langle W_C\, A_{2\ell+1} \rangle \, .
\end{align}
In the next sections we show how we can get the general expressions for the coefficients $\langle W_C\, A_{2\ell+1} \rangle$ valid for any value of the 't Hooft coupling. As an example we start considering the simplest case, i.e. $\ell=1$.

\section{Warm up case: the correlator  \texorpdfstring{$\langle W_C \, O_{3} \rangle$}{}}
\label{sec:example}
Before addressing the most general case, we start considering the simplest correlation function among those we want to study, namely the expectation value of a circular Wilson loop with the insertion of a single-trace chiral operator of dimension $3$. Since the Gram-Schmidt coefficient $M_{3,3} =1$, the expression \eqref{START} for the case at hand reads 
\begin{align}
\langle W_C \, O_3 \rangle = \langle W_C \, A_3 \rangle ~.
\end{align}
In the following we first derive an exact result for this correlator, which is valid for any value of the 't Hooft coupling. Then, exploiting both the asymptotic expressions of the Bessel functions as well as numerical methods, we provide a prediction for its leading term at strong coupling. 

\subsection{Analytical result}

From the definition of the circular Wilson loop in \eqref{WC}, we find
\begin{align}
\label{wp3}
\langle W_C \, O_3 \rangle = \frac{1}{N}\sum_{p=0}^{\infty} \frac{1}{p!}\biggl(\frac{\lambda}{2N}\biggr)^{\frac{p}{2}} \langle A_p\, A_3 \rangle \, ,
\end{align}
where we have replaced the Yang-Mills coupling $g$ with the 't Hooft one $\lambda$. In the equation above, we can actually let the sum start from $p=3$, since the correlator $\langle A_p\, A_3 \rangle $ vanishes for lower values of $p$. Then, knowing that the 2-point function is non-vanishing only if the operators have both odd or even dimension, we can write
\begin{align}
\label{wk3}
\langle W_C \, O_3 \rangle = \frac{1}{N}\sum_{k=1}^{\infty} \frac{1}{(2k+1)!}\biggl(\frac{\lambda}{2N}\biggr)^{k+\frac{1}{2}} \langle A_{2k+1} \, A_3 \rangle \, .
\end{align}
Now we exploit \eqref{eq:AANgrande} to rewrite the 2-point correlator of two odd traces in terms of the coefficients $\textsf{D}_{k,\ell}$ in the planar limit and, after some simple algebraic manipulations, we find
\begin{align}
\label{w3}
\langle W_C \, O_3 \rangle \simeq \frac{\sqrt{\mathcal{G}_3}}{N}\sum_{k=1}^{\infty} \biggl(\frac{\sqrt{\lambda}}{2}\biggr)^{2k+1}\,\sum_{i=0}^{k-1}\frac{\sqrt{2k+1-2i}}{i!(2k+1-i)!}\textsf{D}_{1,k-i} \, ,
\end{align} 
where $\mathcal{G}_3$ is defined in \eqref{Gn}. At this stage we consider the following redefinition of the indices
\begin{align}
\label{redef}
& k-i\longrightarrow n \, ,
\end{align}
therefore \eqref{w3} can be rewritten as
\begin{align}
\label{wcO3}
\langle W_C \, O_3 \rangle \simeq \frac{\sqrt{\mathcal{G}_3}}{N}\sum_{n=1}^{\infty} \sqrt{2n+1}\,\textsf{D}_{1,n}\sum_{i=0}^{\infty} \biggl(\frac{\sqrt{\lambda}}{2}\biggr)^{2i+2n+1}\frac{1}{i!(i+2n+1)!} \, .
\end{align} 
Finally, recalling that the modified Bessel functions of the first kind have the following expansion
\begin{align}
\label{besselI}
I_{\alpha}(x)=\sum_{i=0}^{\infty} \frac{1}{i!(i+\alpha)!} \biggl( \frac{x}{2} \biggr)^{2i+\alpha} \, ,
\end{align}
we can rewrite \eqref{wcO3} as
\begin{align}
\label{WO3}
\langle W_C\,O_3 \rangle \simeq \frac{\sqrt{\mathcal{G}_3}}{N}\sum_{n=1}^{\infty} \sqrt{2n+1}\,\textsf{D}_{1,n}I_{2n+1}(\sqrt{\lambda}) \, .
\end{align}
This expression includes the complete resummation of the perturbative series in the 't Hooft coupling. By exploiting the well-known expansions for small  $\lambda$ of the Bessel functions, contained in the definition of the coefficients $\mathsf{D}_{k,\ell}$, and of the modified Bessel functions, it is possible to generate the perturbative series. On the other hand, making use of the asymptotic expansions of these functions, it also becomes accessible to extrapolate information about the strong coupling behaviour of this correlator. 

It is worth noting that, if we turn off the interaction from the matrix model, then $\mathsf{D}_{1,n}$ becomes $\delta_{1,n}$ and the $\mathcal{N}=4$ result \eqref{N=4} is immediately recovered.

\paragraph{Strong coupling regime}
In order to study the strong coupling behaviour of the correlation function in \eqref{WO3}, we consider its ratio with the corresponding  $\mathcal{N}=4$ correlator, i.e. the function $1+\Delta w_3(\lambda)$ defined in \eqref{target}, namely
\begin{align}
\label{Delta3}
& 1 + \Delta w_3(\lambda) \simeq \frac{1}{\sqrt{3}}\sum_{n=1}^{\infty} \sqrt{2n+1}\,\frac{I_{2n+1}(\sqrt{\lambda})}{I_3(\sqrt{\lambda})}\textsf{D}_{1,n}\, .
\end{align}
Firstly, we recall that a generic modified Bessel function of the first kind $I_{\alpha}(\sqrt{\lambda})$ admits the following asymptotic expansion
\begin{align}
\label{Iexpansion}
 I_{\alpha}(\sqrt{\lambda})  \  \underset{\lambda\rightarrow\infty}{\sim} \ \frac{e^{\sqrt{\lambda}}}{\sqrt{2\pi\lambda^{1/2}}}\left( 1 \ - \  \frac{4\alpha^2-1}{8\sqrt{\lambda}} \ + \  \frac{(4\alpha^2-1)(4\alpha^2-9)}{2!(64\lambda)} \ + \ \dots \right) \, .
\end{align}
Therefore we expand the ratio between the two Bessel functions in \eqref{Delta3} as
\begin{align}
 \frac{I_{2n+1}(\sqrt{\lambda})}{I_{3}(\sqrt{\lambda})} \  \underset{\lambda\rightarrow\infty}{\sim} \  1 - \frac{2n^2+2n-4}{\lambda^{1/2}} + \frac{2n^4+4n^3-7n^2-9n+10}{\lambda} + \dots = \sum_{s=0}^{\infty}\frac{Q_{2s}^{(1)}(n)}{\lambda^{s/2}} \, , 
\end{align}
where $Q^{(1)}_{2s}(n)$ denotes a polynomial in $n$ of degree $2s$. Since we want to study the strong coupling limit of \eqref{Delta3}, we consider the function
\begin{align}
\label{StrongDeltaW3}
\frac{1}{\sqrt{3}}\sum_{n=1}^{\infty}\sqrt{2n+1}\left(\sum_{s=0}^{\infty} \frac{Q^{(1)}_{2s}(n)}{\lambda^{s/2}}\right) \textsf{D}_{1,n} = \frac{1}{\sqrt{3}}\sum_{n=1}^{\infty} \sqrt{2n+1} \textsf{D}_{1,n} + R_3(\lambda)\, , 
\end{align}
where
\begin{align}
\label{R3}
    R_3(\lambda) \equiv \frac{1}{\sqrt{3}}\sum_{n=1}^{\infty}\sqrt{2n+1}\left(\sum_{s=1}^{\infty} \frac{Q^{(1)}_{2s}(n)}{\lambda^{s/2}}\right) \textsf{D}_{1,n}\, .
\end{align}
In the first term on the r.h.s of \eqref{StrongDeltaW3} we recognize the expansion of the coefficient $\mathsf{d}_1$ defined in \eqref{dk}, whose leading order behaviour at strong coupling has been analytically obtained in \cite{Billo:2022fnb} and reported in \eqref{dstrong}. Hence we conclude that 
\begin{align}
\label{finalleadingW3}
1 + \Delta w_3(\lambda) \underset{\lambda\rightarrow\infty}{\sim} \biggl(\frac{4\pi}{\sqrt{\lambda}}+O\biggl( \frac{1}{\lambda} \biggr)\biggr) + R_3(\lambda) \, ,
\end{align}
where we still have to determine the leading term of the large-$\lambda$ expansion of $R_3(\lambda)$.  To this regard we observe that the $s$th term in \eqref{R3} is suppressed by the $s$th power of the square root of the coupling and therefore this could lead to the conclusion that $R_3(\lambda)$ does not contribute (at least to the leading order of the expansion). On the another hand this effect could be compensated by the increasing degree of the polynomials $Q^{(1)}_{2s}(n)$\footnote{We are indebted to G.P. Korchemsky for drawing our attention to this aspect.}. Although we do not have an analytical way to determine the strong coupling contribution of $R_3(\lambda)$, we can employ numerical methods to clarify this aspect. This analysis will be carried out in next section.

\subsection{Numerical analysis}
Here we clarify if the term $R_3(\lambda)$ \eqref{R3} contributes to the leading order of $1+\Delta w_3(\lambda)$ using numerical methods. As a first step, using the perturbative expansion for the $\textsf{X}$-matrix \eqref{Xmatrix} and its powers \eqref{X2}, we generate very long perturbative series for the coefficients $\mathsf{D}_{1,\ell}$ \eqref{D}. For example the first orders of the series expansion for  $\mathsf{D}_{1,1}$  can be obtained starting from \eqref{X11pert}-\eqref{X311pert} and read
\begin{align}
\mathsf{D}_{1,1}(\lambda) = 1 - \frac{5\,\zeta_5}{256\pi^6}\lambda^3 + \frac{105\,\zeta_7}{4096\pi^8}\lambda^4 - \frac{1701\,\zeta_9}{65536\pi^{10}}\lambda^{5}+\left(\frac{25\,\zeta_5^2}{65536\pi^{12}}+\frac{12705\,\zeta_{11}}{524288\pi^{12}}\right)\lambda^6 + \dots\, ,
\end{align}
where the dots stand for higher orders of the expansion. In general the series expansion for the coefficient $\mathsf{D}_{1,\ell}$ (with $\ell>1$) begins with a term proportional to $\lambda^{\ell+2}$. Then, using these expressions as well as the relation \eqref{Delta3}, the series expansion for $1 + \Delta w_{3}(\lambda)$ can be obtained as
\begin{align}
\label{Delta3Series}
1 + \Delta w_3(\lambda) =   \frac{1}{\sqrt{3}}\sum_{n=1}^{M} \sqrt{2n+1}\,\textsf{D}_{1,n}(\lambda)\frac{I_{2n+1}(\sqrt{\lambda})}{I_3(\sqrt{\lambda})} \equiv 1 + \sum_{k=3}^{M} \Delta w_3^{(k)}\left(\frac{\lambda}{\pi^2}\right)^k \, , 
\end{align}
where the sum stops at some integer $M$. We observe that the truncation of the above sum at $M$ implies that we need to compute the series expansions for the first $\mathsf{D}_{1,M-2}$ up to the order $\lambda^M$. In order to find a good balance among the computational cost and the need to generate enough precise numerical results we chose $M=100$. 

\begin{figure}%
    \centering
    {{\includegraphics[scale=0.23]{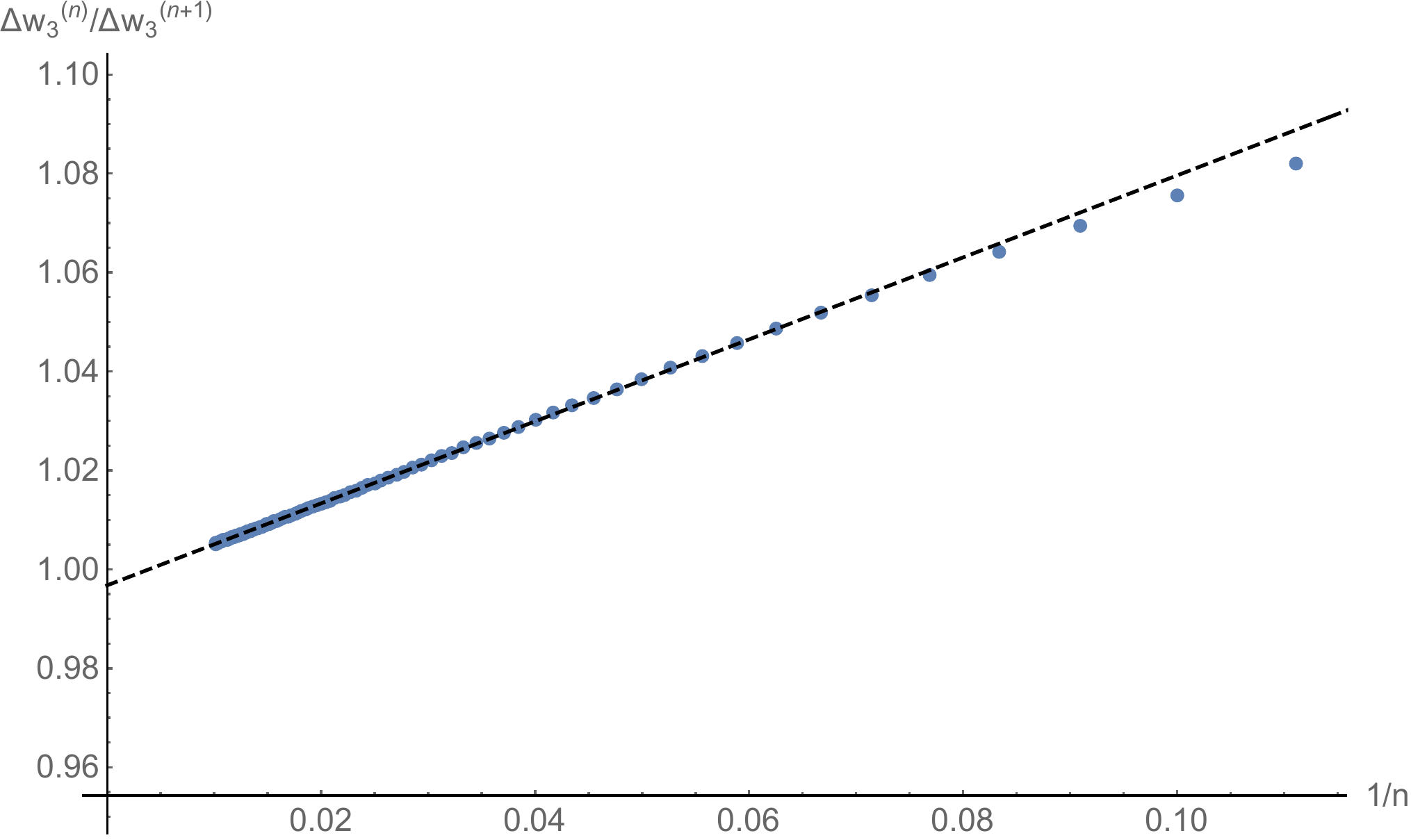} }}%
    {{\includegraphics[scale=0.40]{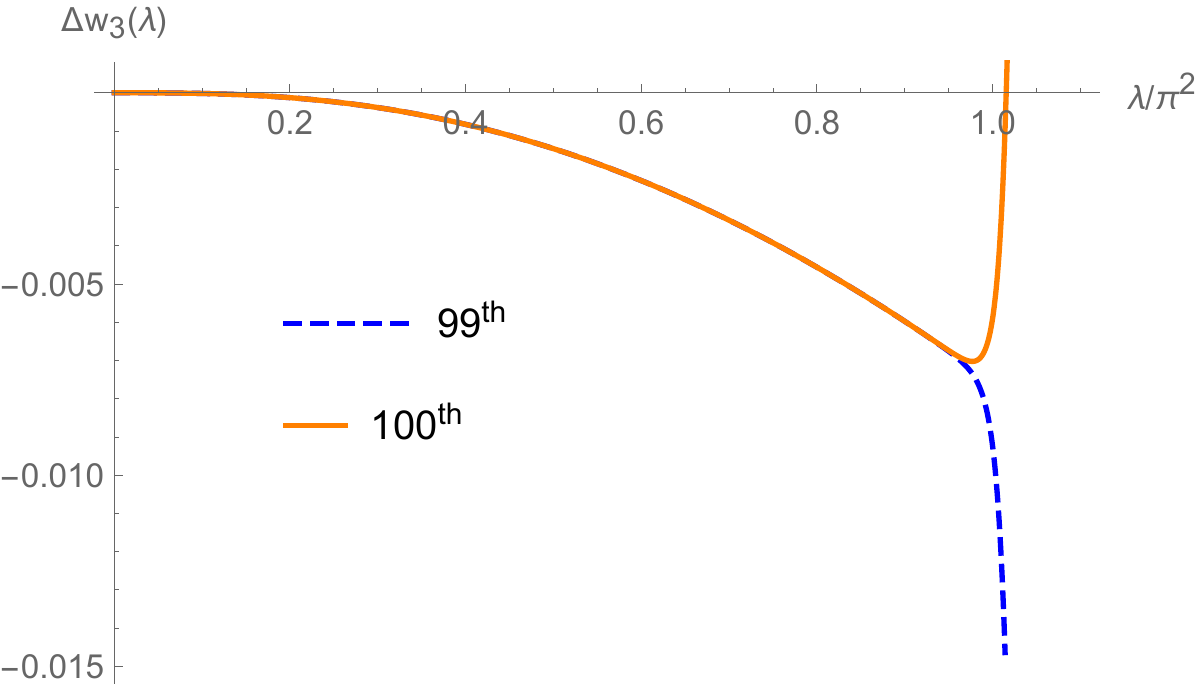} }}%
    \caption{On the left we reported the plot for the estimation of the radius of convergence of the series \eqref{Delta3Series}. The intercept of the linear fit is $0.996$. On the right we reported the 99-th and 100-th order truncated series of $\Delta w_3(\lambda)$. }%
    \label{fig:Radius}%
\end{figure}

\begin{figure}[ht!]
\centering
\centering
    {{\includegraphics[scale=0.19]{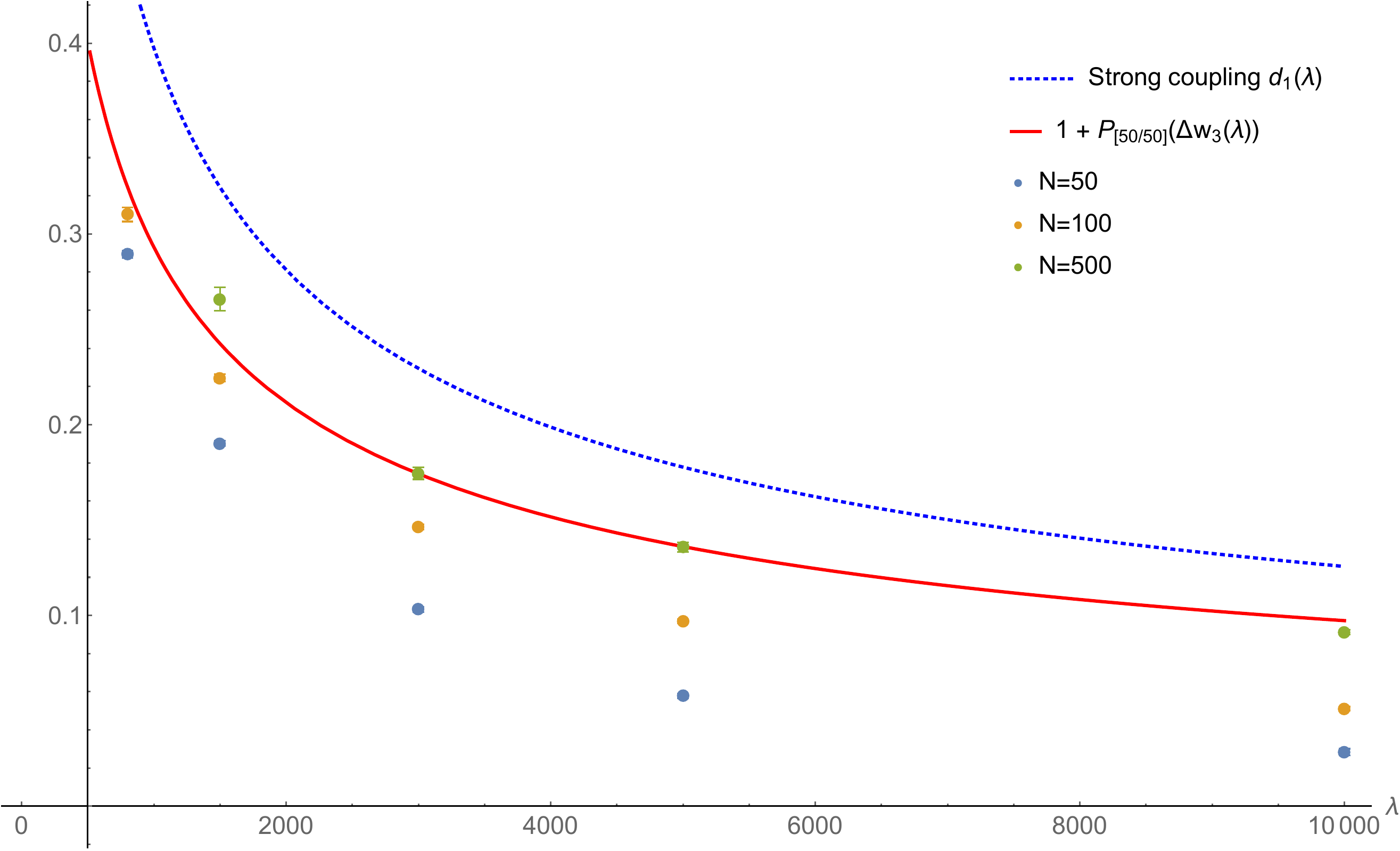} }}%
    {{\includegraphics[scale=0.18]{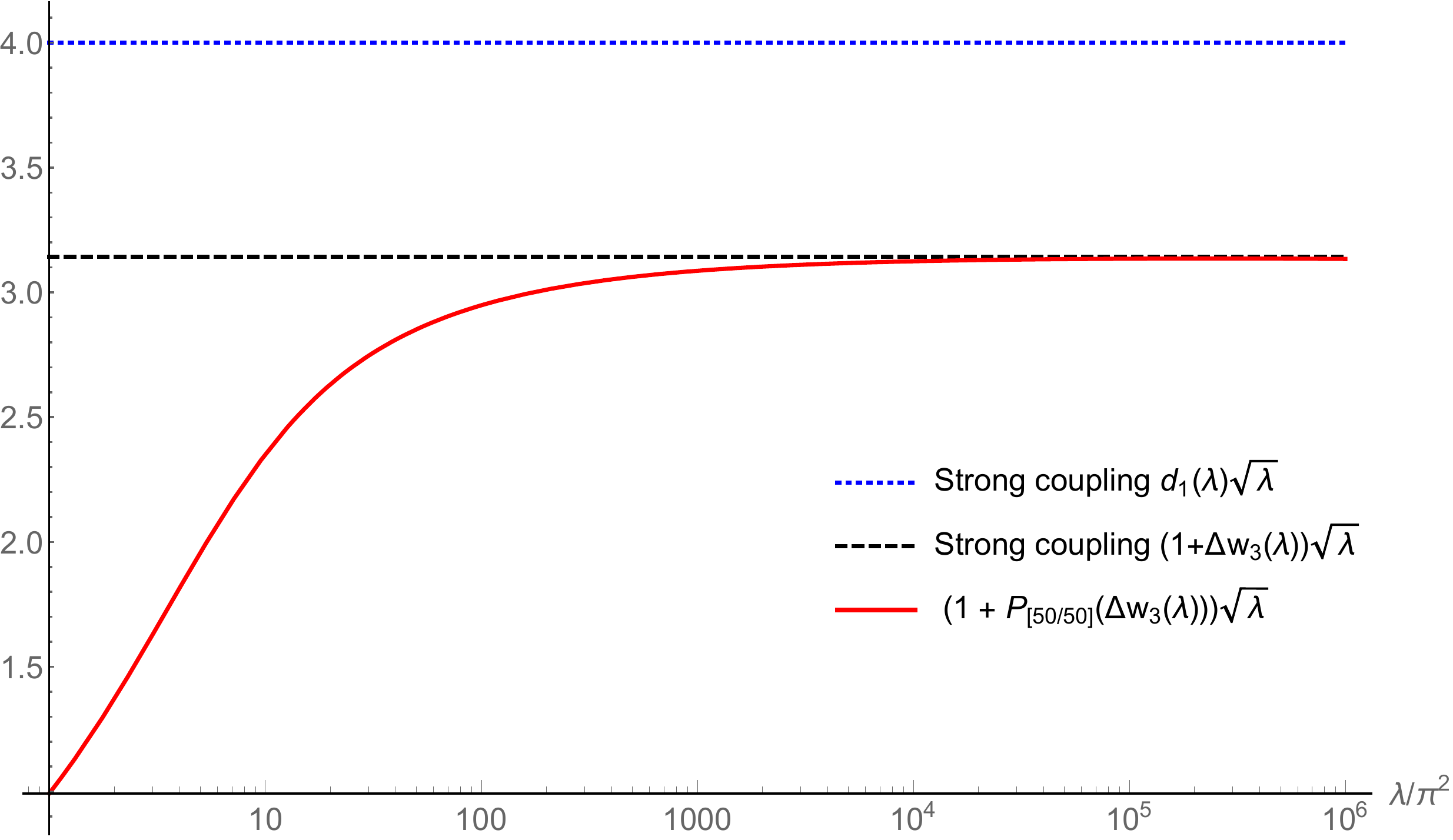} }}%
\caption{On the left we reported the comparison among  the large $\lambda$ theoretical prediction for $\textsf{d}_1$ \eqref{dk} (blue dotted line), the Padé curve for $q=50$ for the function $1+\Delta w_3(\lambda)$ (red curve) and the points from the Monte Carlo simulations at $N=50$ (blue circles), at $N=100$ (orange circles) and at $N=500$ (green circles). On the right we reported the comparison among the leading term of $\textsf{d}_1(\lambda)\sqrt{\lambda}$ (blue dotted line) and the conformal-Padé (red curve) multiplied by $\sqrt{\lambda}$. It is evident the presence of a finite gap between the blue curve and red curve. Furthermore we reported the strong coupling prediction \eqref{Deltaw3Final} for $(1+\Delta w_3(\lambda))$ multiplied  by $\sqrt{\lambda}$ (black dashed line). We observe that the red curve and the black curve tend to the same value for large values of the 't Hooft coupling.}
\label{fig:MCandPadeWO3}
\end{figure}

Then using the ratio test we estimate the radius of convergence $L$ of the series \eqref{Delta3Series}, namely
\begin{align}
    L = \lim_{n \to \infty} \Big| \frac{\Delta w_3^{(n)}}{\Delta w_3^{(n+1)}} \Big| \, .
\end{align}
The result is reported on the left hand side of Fig. \ref{fig:Radius}. According to our analysis the series \eqref{Delta3Series} has finite radius of convergence located at $\lambda \simeq \pi^2$. This expectation is confirmed by the plot of the 99-th and 100-th truncated series for $\Delta w_3(\lambda)$ (see the right hand side of Fig. \ref{fig:Radius}).
We observe an alternating numerical blow up near the value $\lambda/\pi^2 \simeq 1$. The series \eqref{Delta3Series} can be extended beyond its radius of convergence via a Padé resummation. As it has been done in \cite{Beccaria:2021hvt,Billo:2022xas}  a suitable choice for the Padé approximant is to consider the diagonal ones
\begin{align}
\label{PadeGeneral}
P_{[q/q]}(\Delta w_3) = \left[\sum_{k=3}^{100}\Delta w_3^{(k)}\left(\frac{\lambda}{\pi^2}\right)^k \right]_{[q/q]},
\end{align}
where $q$ is an integer. Furthermore, before computing the Padé approximant, it is very convenient to perform a conformal map. Following \cite{Beccaria:2021hvt} we make the replacement
\begin{align}
    \frac{\lambda}{\pi^2} \mapsto \frac{4z}{(z-1)^2}\, ,
\end{align}
then we construct the Padé approximant in the variable $z$ inside the unit circle $|z| \leq 1$ and, as a last step, we replace back $z$ with $\lambda$ using the map
\begin{align}
    z = \frac{\sqrt{1+\frac{\lambda}{\pi^2}}-1}{\sqrt{1+\frac{\lambda}{\pi^2}}+1} \, .
\end{align}
As it was discussed in \cite{Costin:2019xql,Costin:2020hwg,Beccaria:2021vuc} one of the main advantage of this procedure is that inside $|z| \leq 1$ the expansion is convergent by construction. This in turn permits to obtain a conformal-Padé approximant that is stable even for very large values of the coupling (see the right hand side of Fig. \ref{fig:MCandPadeWO3}).

Finally, as a further independent numerical check, we choose to perform a Monte Carlo simulation of the function $1 + \Delta w_3(\lambda)$ using a Metropolis-Hastings algorithm \cite{brooks2011handbook} following the same procedure discussed in \cite{Beccaria:2021hvt} and with $N=50,\,100$ and $500$. The results are reported on the left hand side of Fig. \ref{fig:MCandPadeWO3} together with the diagonal Padé approximant with $q=50$ and the strong coupling behaviour of $\mathsf{d}_1$ \eqref{dk}. We observe that, as the values of $N$ increase, the Monte Carlo points approach the Padé curve. In particular the points with $N=500$ lie very close to the Padé curve, strongly suggesting that the two independent numerical methods agree with each other. Furthermore we observe that, for larger values of $\lambda$, the Padé curve seems to tend towards the strong coupling behaviour of $\mathsf{d}_1$. Nevertheless, based only on these data, we cannot draw any conclusions concerning a possible contribution due to $R_3(\lambda)$. To clarify this point we chose to multiply both the conformal-Padé  and the strong coupling behaviour of $\textsf{d}_1$ by $\sqrt{\lambda}$ and then numerically compare the two curves. The corresponding result is reported on the right hand side of Fig. \ref{fig:MCandPadeWO3}, where we observe the presence of a finite (almost constant) gap between the two lines. This in turn strongly suggests that the strong coupling expansion of $R_3(\lambda)$ should start with a term proportional to $\lambda^{-1/2}$.

Finally we numerically evaluate the difference between the conformal-Padé and the strong coupling behaviour of the $\textsf{d}_1$ coefficient. We choose to perform this analysis in the region $10^3 \leq \lambda/\pi^2 \leq 10^6$ (where the conformal-Padé is almost constant) and we perform a fit of the difference, namely the function $R_3(\lambda)$, using the following ansatz
\begin{align}
 R_3(\lambda) = \sum_{j=0}^{P}\frac{R_3^{(j)}}{\lambda^{j/2}} \, ,
\end{align}
with $P \geq 2$. The outcome of this analysis is that the numerical estimation $R_3^{(0)}$ and $R_3^{(1)}$ (that are the only coefficients of interest for us) is not affected in appreciable way by the change of $P$. We find 
\begin{align}
\label{ccW3}
    R_3^{(0)} \simeq -2.1\,\times\,10^{-5}\, , \ \ \ \ R_3^{(1)}\simeq -2.69(4)\, .
\end{align}
As expected the coefficient $R_3^{(0)}$ is very close to zero. Therefore our conclusion is that the function $R_3(\lambda)$ contributes to the leading order of the strong coupling expansion of \eqref{finalleadingW3} and the numerical estimation of its leading term (for $\lambda >>1$) is
\begin{align}
\label{R3num}
    R_3(\lambda) \simeq \frac{R_3^{(1)}}{\sqrt{\lambda}} + O\left(\frac{1}{\lambda}\right) \, .
\end{align}
Thus, our final expression for the leading term of \eqref{finalleadingW3} reads
\begin{align}
\label{Deltaw3Final}
    1+\Delta w_3(\lambda) \underset{\lambda\rightarrow\infty}{\sim} \frac{1}{\sqrt{\lambda}}\left(4\pi + R_3^{(1)}\right) + O\biggl( \frac{1}{\lambda} \biggr) \, .
\end{align}

\section{The general case \texorpdfstring{$\langle W_C \,O_{2p+1}\rangle$}{}}
\label{sec:general}
In this section we consider the general case. We initially find an analytical expression which resums the entire perturbative series in the planar limit for the correlator $\langle W_C\,A_{2\ell+1} \rangle$ and, then, we exploit this exact formula to study the gauge theory correlator $\langle W_C\,O_{2p+1} \rangle$ at the leading order at strong coupling. 

The correlation function of interest reads
\begin{align}
\label{wkl}
\langle W_C \, A_{2\ell+1} \rangle = \frac{1}{N}\sum_{k=1}^{\infty} \frac{1}{(2k+1)!}\biggl(\frac{\lambda}{2N}\biggr)^{k+\frac{1}{2}} \langle A_{2k+1}\, A_{2\ell+1} \rangle \, .
\end{align}
Using \eqref{eq:AANgrande} and doing some simple algebra, one immediately gets
\begin{align}
\label{wl}
\langle W_C \, A_{2\ell+1} \rangle \simeq \frac{\sqrt{\mathcal{G}_{2\ell+1}}}{N}\sum_{k=1}^{\infty} \biggl(\frac{\sqrt{\lambda}}{2}\biggr)^{2k+1}\,\sum_{i=0}^{\infty}\,\sum_{j=0}^{\ell-1}\frac{\sqrt{2k+1-2i}\sqrt{2\ell+1-2j}}{i!(2k+1-i)!\sqrt{2\ell+1}}\binom{2\ell+1}{j}\textsf{D}_{k-i,\ell-j} \, .
\end{align} 
At this point we redefine the sum indices as follows
\begin{align}
& k-i\longrightarrow n\, , \ \ \ \ \ \ell-j\longrightarrow m  \label{nindex}   
\end{align}
and, using again the series expansion of the modified Bessel function in \eqref{besselI}, the expression \eqref{wl} becomes
\begin{align}
\label{finalwl}
\langle W_C \, A_{2\ell+1} \rangle \simeq \frac{\sqrt{\mathcal{G}_{2\ell+1}}}{N}\sum_{n=1}^{\infty} \sqrt{2n+1} \left( I_{2n+1}(\sqrt{\lambda})\,\sum_{m=1}^{\ell}\textsf{h}_m^{(\ell)}\,\textsf{D}_{n,m} \right) \, ,
\end{align} 
where the numerical coefficients $\textsf{h}_m^{(\ell)}$ read
\begin{align}
\label{h}
\textsf{h}_m^{(\ell)}=\sqrt{\frac{2m+1}{2\ell+1}}\binom{2\ell+1}{\ell-m} \, .
\end{align}
Inserting \eqref{finalwl} in \eqref{START}, we obtain
\begin{align}
\label{FullResult}
\langle W_C\,O_{2p+1} \rangle \simeq \frac{1}{N}\sum_{\ell=1}^{p}\sqrt{\mathcal{G}_{2\ell+1}}M_{2p+1,2\ell+1}(\lambda)\sum_{n=1}^{\infty} \sqrt{2n+1} \left( I_{2n+1}(\sqrt{\lambda})\,\sum_{m=1}^{\ell}\textsf{h}_m^{(\ell)}\,\textsf{D}_{n,m} \right) \, . 
\end{align}
This is our final result for the correlator $\langle W_C\, O_{2p+1} \rangle$, which can be used both to generate very long perturbative series in $\lambda$ or to investigate the strong coupling regime of $1+ \Delta w_{2p+1}(\lambda)$. 

\subsection{Strong coupling regime}
 Using \eqref{FullResult} we determine the general expression for the function \eqref{START}
\begin{align}
\label{start1}
1+\Delta w_{2p+1}(\lambda) = \frac{1}{\sqrt{(2p+1)\mathcal{G}_{2p+1}}}\sum_{\ell=1}^{p}\sqrt{\mathcal{G}_{2\ell+1}}M_{2p+1,2\ell+1}(\lambda)\sum_{n=1}^{\infty} \sqrt{2n+1} \left( \frac{I_{2n+1}(\sqrt{\lambda})}{I_{2p+1}(\sqrt{\lambda})}\,\sum_{m=1}^{\ell}\textsf{h}_m^{(\ell)}\,\textsf{D}_{n,m} \right) \, .
\end{align}
We now move to consider the strong coupling limit of \eqref{start1}.
The first step is to observe that at the leading order of the strong coupling expansion the expression for the Gram-Schmidt coefficients $M_{2p+1,2\ell+1}(\lambda)$ drastically simplifies and it does not depend on $\lambda$ \cite{Billo:2022xas}
\begin{align}
\label{Mstrong}
\lim_{\lambda \to \infty} M_{2p+1,2\ell+1}(\lambda) \equiv M^{(\infty)}_{2p+1,2\ell+1} = \biggl(-\frac{N}{2}\biggr)^{p-\ell}\, \frac{2p+1}{2\ell+1} \frac{p}{\ell}\, \binom{p+\ell-1 }{p-\ell}\, .
\end{align}
We adopt the same procedure used for the $p=1$ case discussed in Section \ref{sec:example}. Namely we expand the ratio between the two Bessel functions in \eqref{start1} using \eqref{Iexpansion}. Since we are interested in the leading term of the strong coupling expansion of \eqref{start1} we consider the expression
\begin{align}
\label{W2p1}
\frac{1}{\sqrt{(2p+1)\mathcal{G}_{2p+1}}}\sum_{\ell=1}^{p}\sqrt{\mathcal{G}_{2\ell+1}}M^{(\infty)}_{2p+1,2\ell+1}\sum_{n=1}^{\infty}\sqrt{2n+1}\left(\sum_{s=0}^{\infty}\frac{Q_{2s}^{(p)}(n)}{\lambda^{s/2}} \sum_{m=1}^{\ell}\textsf{h}_m^{(\ell)}\,\textsf{D}_{n,m}\right)\, ,    
\end{align}
where $Q_{2s}^{(p)}(n)$ denotes a polynomial in $n$ of degree $2s$. For example for $s=0,1$  we find
\begin{align}
\label{Qpoly}
Q_{0}^{(p)}(n) =1, \ \ \ \ \ Q_{2}^{(p)}(n) = 2(p-n)(1+p+n)\, . 
\end{align}
Then we separate the contribution due to $Q_0^{(p)}(n)$ and that of other polynomials $Q^{(p)}_{2s}(n)$ with $s \geq 1$. Therefore we rewrite \eqref{W2p1} as
\begin{align}
\label{start2}
\frac{1}{\sqrt{(2p+1)\mathcal{G}_{2p+1}}}\sum_{\ell=1}^{p}\sqrt{\mathcal{G}_{2\ell+1}}M^{(\infty)}_{2p+1,2\ell+1}\sum_{m=1}^{\ell}\textsf{h}_m^{(\ell)}\left(\sum_{n=1}^{\infty}\sqrt{2n+1}\textsf{D}_{n,m}\right) +R_{2p+1}(\lambda) \, ,
\end{align}
where
\begin{align}
\label{R2p1}
R_{2p+1}(\lambda) \equiv \frac{1}{\sqrt{(2p+1)\mathcal{G}_{2p+1}}}\sum_{\ell=1}^{p}\sqrt{\mathcal{G}_{2\ell+1}}M^{(\infty)}_{2p+1,2\ell+1}\sum_{n=1}^{\infty}\sqrt{2n+1}\left(\sum_{s=1}^{\infty}\frac{Q_{2s}^{(p)}(n)}{\lambda^{s/2}} \sum_{m=1}^{\ell}\textsf{h}_m^{(\ell)}\,\textsf{D}_{n,m}\right)  \, .  
\end{align}
Inside the bracket in \eqref{start2} we recognise the expansion of the coefficient $\mathsf{d}_m$  \eqref{dk}, whose leading order behaviour at strong coupling has been reported in \eqref{dstrong}. Furthermore in Appendix \ref{appendix:A} we argue that at leading order in the large-$\lambda$ expansion it holds that
\begin{align}
\label{largeR2p1}
R_{2p+1}(\lambda) \underset{\lambda\rightarrow\infty}{\sim} \frac{R_3^{(1)}}{\sqrt{(2p+1)\mathcal{G}_{2p+1}}}\sum_{\ell=1}^{p}\sqrt{\mathcal{G}_{2\ell+1}}M^{(\infty)}_{2p+1,2\ell+1}\sum_{m=1}^{\ell}\textsf{h}^{(\ell)}_{m}\left(\frac{\sqrt{2m+1}(m^2+m)}{2\,\sqrt{\lambda}}\right) + O\left(\frac{1}{\lambda}\right) \, ,
\end{align}
where the coefficient $R_3^{(1)}$ was evaluated numerically in \eqref{ccW3}. Then inserting \eqref{largeR2p1} in \eqref{start2}  
 and using the expressions \eqref{Mstrong}, \eqref{h} and \eqref{dstrong}, in the large-$\lambda$ limit we find
\begin{align}
\label{leadingWl2}
 1 + \Delta w_{2p+1}(\lambda) \underset{\lambda\rightarrow\infty}{\sim}  
 \frac{1}{\sqrt{\lambda}}\left(2\pi+\frac{R_3^{(1)}}{2}\right)2p\,(-1)^{p}\sum_{\ell=1}^{p}\,\frac{(-1)^{\ell}(p+\ell-1)!}{(p-\ell)!}\sum_{m=1}^{\ell}\frac{(2m+1)(m^2+m)}{(\ell-m)!(\ell+m+1)!} + O\left(\frac{1}{\lambda}\right) \, .
\end{align}
We perform the sum over $m$ using the following identity
\begin{align}
    \sum_{m=1}^{\ell}\frac{(2m+1)(m^2+m)}{(\ell-m)!(\ell+m+1)!} = \frac{1}{(\ell-1)!\ell!} \, \ ,
\end{align}
this way the expression \eqref{leadingWl2} simplifies and becomes
\begin{align}
\label{stepint}
 1 + \Delta w_{2p+1}(\lambda) \underset{\lambda\rightarrow\infty}{\sim}  
 \frac{1}{\sqrt{\lambda}}\left(2\pi+\frac{R_3^{(1)}}{2}\right)2p(-1)^{p}\sum_{\ell=1}^{p}\,\frac{(-1)^{\ell}(p+\ell-1)!}{(p-\ell)!(\ell-1)!\ell!}\, .   
\end{align}
As a final step we notice that the also the sum over $\ell$ can be done analytically using the mathematical identity
\begin{align}
    \sum_{\ell=1}^{p}\frac{(-1)^{\ell}(p+\ell-1)!}{(p-\ell)!(\ell-1)!\ell!}   = (-1)^p \, .
\end{align}
Inserting this back in \eqref{stepint} we obtain the following prediction for the leading term of the strong coupling expansion of $1+\Delta w_{2p+1}(\lambda)$
\begin{align}
\label{finalleadingWl}
1 + \Delta w_{2p+1}(\lambda) \underset{\lambda\rightarrow\infty}{\sim} \frac{2p}{\sqrt{\lambda}}\left(2\pi+\frac{R_3^{(1)}}{2}\right) + O\left(\frac{1}{\lambda}\right) \, .
\end{align}
This is our final result for the leading term of the strong coupling expansion of \eqref{start1}. 

\begin{figure}[ht!]
\centering
\centering
    {{\includegraphics[scale=0.16]{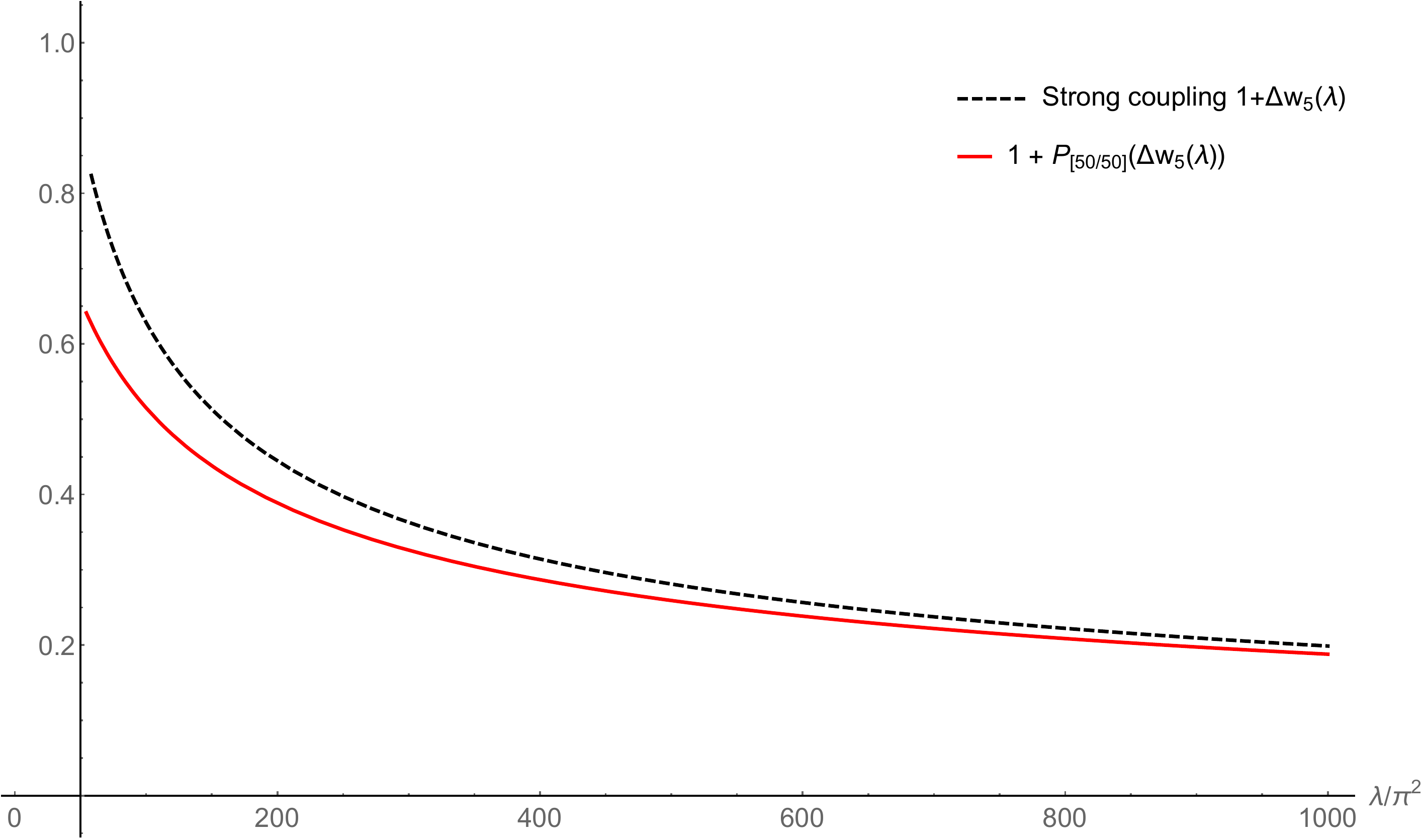} }}%
    {{\includegraphics[scale=0.22]{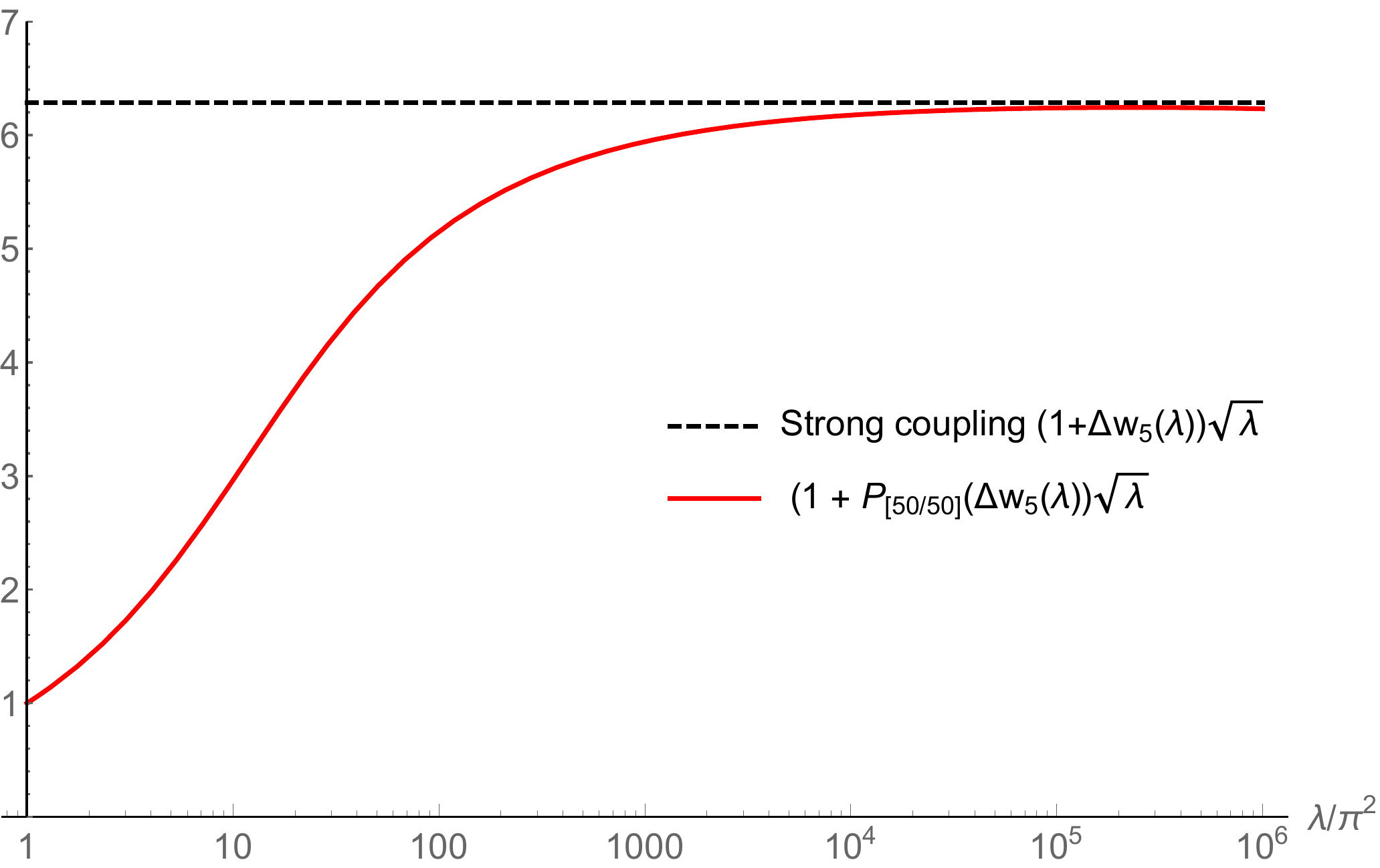} }}%
\caption{On the left we reported the comparison between  the large $\lambda$ theoretical prediction for $1+\Delta w_5(\lambda)$ \eqref{finalleadingWl} (black dashed line) and the Padé curve for $q=50$ for the function $\Delta w_5(\lambda)$ (red curve). On the right we reported the comparison between $1+\Delta w_5(\lambda)$ (black dashed line) and the conformal-Padé (red curve), both functions have been multiplied by $\sqrt{\lambda}$.}
\label{fig:WO5}
\end{figure}

\begin{figure}[ht!]
\centering
\centering
    {{\includegraphics[scale=0.17]{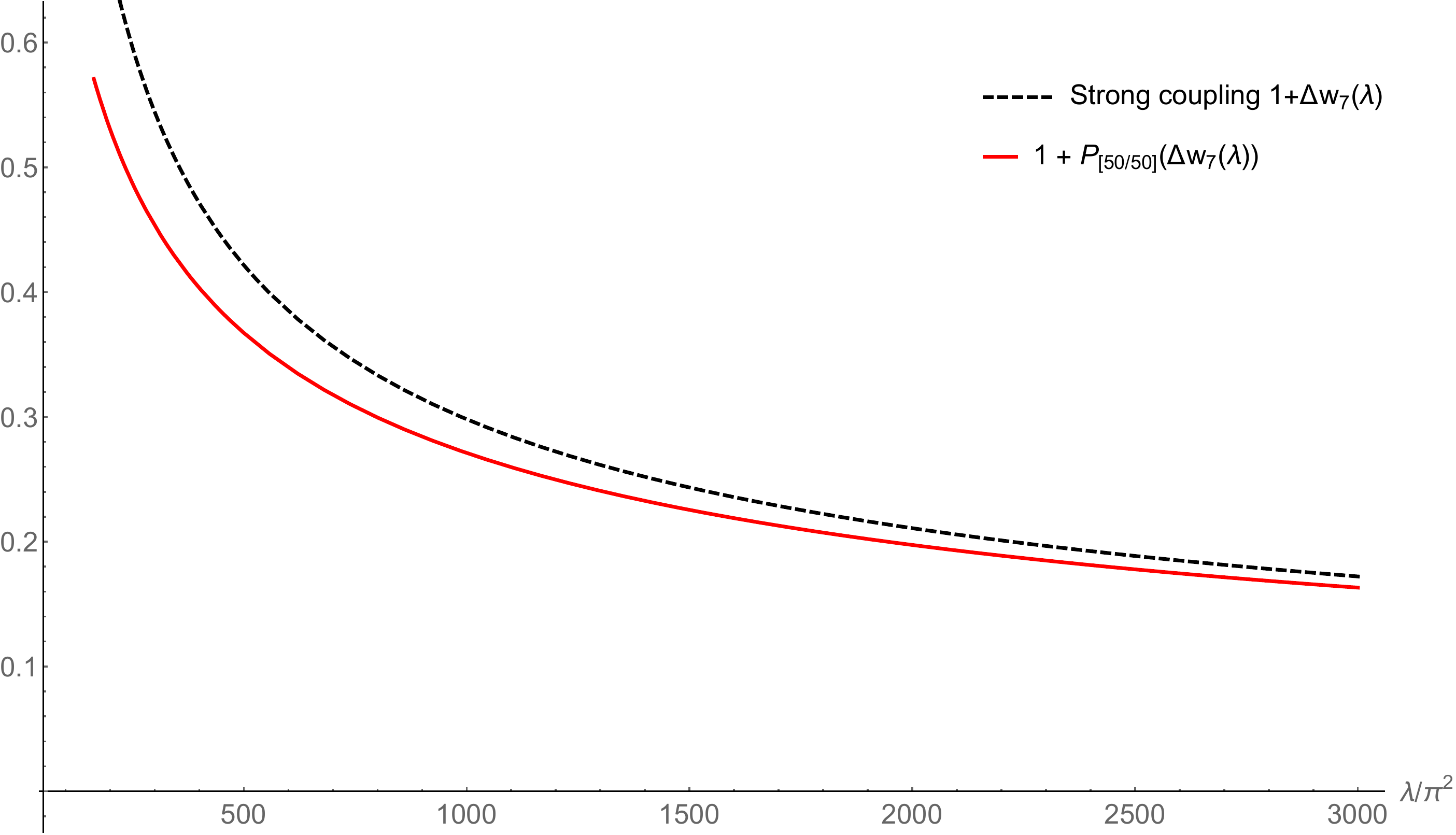} }}%
    {{\includegraphics[scale=0.17]{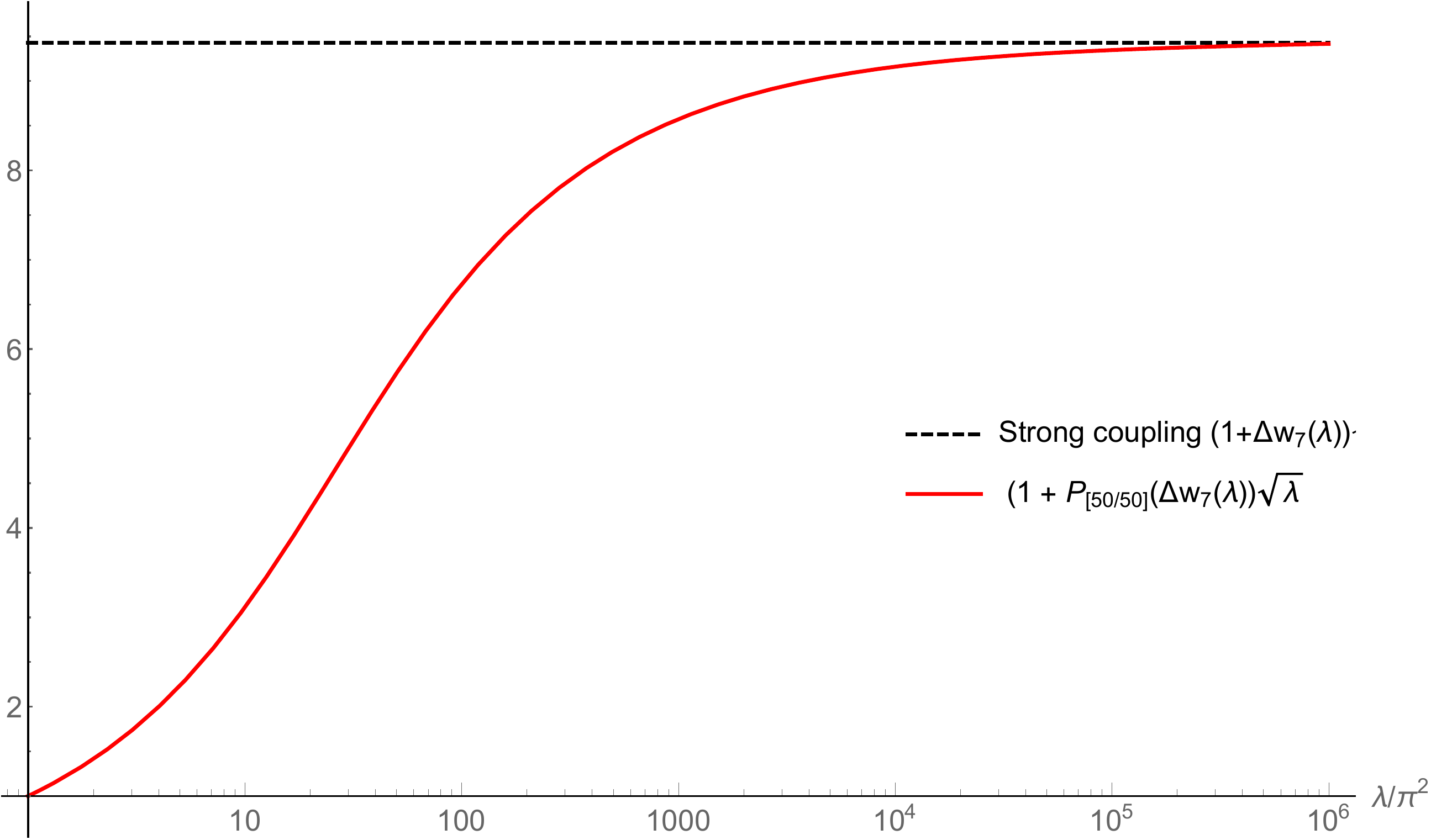} }}%
\caption{On the left we reported the comparison between  the large $\lambda$ theoretical prediction for $1+\Delta w_7(\lambda)$ \eqref{finalleadingWl} (black dashed line) and the Padé curve for $q=50$ for the function $\Delta w_7(\lambda)$ (red curve). On the right we reported the comparison between $1+\Delta w_7(\lambda)$ (black dashed line) and the conformal-Padé (red curve), both functions have been multiplied by $\sqrt{\lambda}$.}
\label{fig:WO7}
\end{figure}

Using the same method discussed in Section \ref{sec:example}, we can provide some numerical checks for the general strong coupling prediction \eqref{finalleadingWl}, namely we numerically extend the perturbative series for \eqref{start1} beyond  its radius of convergence (located at $\lambda \simeq \pi^2$) through a conformal-Padé resummation. The results for the $p=2$ and the $p=3$  cases are shown in Fig. \ref{fig:WO5} and Fig. \ref{fig:WO7} respectively. We observe that in both cases the conformal-Padé approaches the corresponding large $\lambda$ prediction. We regard these numerical evaluations as a strong confirmation of our prediction \eqref{finalleadingWl}.

\section{Conclusions}
\label{sec:conclusions}

The main result of this paper was to provide an exact expression (the relation \eqref{FullResult}), valid for any value of the 't Hooft coupling in the planar limit, for the correlator among a circular Wilson loop and a chiral primary operator in the \textbf{E}-theory. On the top of that we furnished a prediction for the leading term of its strong coupling large-$\lambda$ expansion, which can be rewritten in terms of the analogous correlator of the $\mathcal{N}=4$ SYM theory in a remarkable simple way
\begin{align}
\label{2p}
 \langle W_C\, O_{2p+1} \rangle \underset{\lambda\rightarrow\infty}{\sim} \langle W_C\, O_{2p+1} \rangle_0 \, \frac{(\Delta_p-1)}{\sqrt{\lambda}}\left(2\pi+\frac{R_3^{(1)}}{2}\right) + O\left(\frac{1}{\lambda}\right) \, ,
\end{align}
where $\Delta_{p} \equiv 2p+1$ denotes the conformal dimension of the chiral operator $O_{2p+1}$ and the numerical coefficient $R_3^{(1)}=-2.69(4)$ has been evaluated in Section \ref{sec:example} \footnote{In principle one could employ the numerical procedure described in Section \ref{sec:example} to estimate also the subleading terms of the large $\lambda$ expansion of $R_{2p+1}(\lambda)$. For example, in the case of $\lambda^{-1}$ coefficient of $R_3(\lambda)$, we found
\begin{align}
R_3^{(2)} \, \simeq \, 35.(5)  \, \ .
\end{align}
However it is important to notice that both the computational cost and the numerical errors grow very quickly with the order of the expansion. Due to these reasons the numerical evaluation of the large $\lambda$ expansion of \eqref{2p} turns out to be a very useful tool only for the leading order of the expansion.} 
. In the future we would like to clarify if the analytical method developed in \cite{Belitsky:2020qrm,Belitsky:2020qir} can be applied also for the computation of this coefficient, in order to provide a non-trivial check of our numerical analysis.

It is very interesting to notice that the same dependence on the conformal dimension was previously found for different correlation functions in \cite{Billo:2022xas,Billo:2022fnb}. There, it was shown that the leading term of the  strong coupling expansion of the 3-point extremal correlator among twisted chiral primary operators $T_{\alpha,k_i}$ of conformal dimension $k_1$, $k_2$ and $k_1+k_2$ factorizes as the product between the $\mathcal{N}=4$ extremal correlator and three factors $\frac{2\pi}{\sqrt{\lambda}}(\Delta_{T_{\alpha,k_i}}-1)$ associated to the chiral operators, namely
\begin{align}
    \langle T_{\alpha,k_1}\  T_{\beta,k_2} \  \overline{T}_{\alpha+\beta,k_1+k_2} \rangle \underset{\lambda\rightarrow\infty}{\sim} \langle T_{\alpha,k_1} \ T_{\beta,k_2} \ \overline{T}_{\alpha+\beta,k_1+k_2} \rangle_0 \prod_{i=1}^{3}\frac{2\pi}{\sqrt{\lambda}}(\Delta_{T_{k_i}}-1)\, .
\end{align}
In the future it would be interesting to further investigate this feature and understand if it holds for different observables, such as correlation functions among one chiral operator and a circular Wilson loop in different representations of the gauge group or to correlators among a Wilson loop and two chiral scalar operators, which have been considered in \cite{Giombi:2012ep,Beccaria:2020ykg} for $\mathcal{N}=4$ SYM. 

Furthermore it would be interesting to study the same defect correlation function in the context of the $4d$ $\mathcal{N}=2$ quiver gauge theory arising as a $\mathbb{Z}_M$ orbifold of $\mathcal{N}=4$ SYM, this way extending the perturbative analysis initiated in \cite{Galvagno:2021bbj}. Moreover, as it was shown in \cite{Billo:2022fnb}, in the case of structure constants of chiral primary operators, the holographic dual geometry
of this circular quiver gauge theory is known and simple enough to allow explicit computations at the Supergravity level. Therefore, using 
the AdS/CFT correspondence, it could also be possible to cross check the analogous of the strong coupling prediction \eqref{finalleadingWl} valid for the circular quiver gauge theory. 

Finally, since the \textbf{E}-theory admits a gravity dual \cite{Ennes:2000fu}, in principle it should be possible to cross check the prediction \eqref{2p} using holography. According to the AdS/CFT dictionary the expectation value of a circular Wilson loop in the fundamental representation is captured by the area of the minimal surface extending in the bulk and ending on the AdS boundary \cite{Maldacena:1998im}. Here we are interested in the v.e.v. of a circular Wilson loop with the insertion of a scalar chiral operator $O_{\Delta}$. The holographic counterpart of this observable has been widely analysed for  $\mathcal{N}=4$ SYM (see for instance \cite{Berenstein:1998ij,Giombi:2006de}). From an holographic point of view the crucial quantity that allows to do the computation is the vertex operator $V_{\Delta}$ encoding the coupling to the string
worldsheet of the supergravity mode dual to the operator $O_{\Delta}$. To the best of our knowledge, in all the cases known in the literature, the expression of the vertex operator $V_{\Delta}$ has obtained starting with the string Nambu-Goto action and expanding it to linear order in the fluctuations of the metric. However, it is important to note that for the \textbf{E}-theory \cite{Ennes:2000fu} only the supergravity modes dual to the (even-dimensional) untwisted chiral operators have a non trivial coupling with the fluctuations of the metric, while the modes dual to the (odd-dimensional) twisted operators, such as the ones considered in this article, have not. Therefore we conclude that, in the present case, the vertex operator $V_{\Delta}$ cannot be obtained using only the Nambu-Goto action. 
The extension of the holographic dictionary to this class of observables is not an immediate generalization of previous results and, therefore, is beyond the scope of the present paper. We plan to further analyse it in the future.

\vskip 1cm
\noindent {\large {\bf Acknowledgments}}
\vskip 0.2cm
We are very grateful to A. Lerda, M. Frau and M. Billò for many important discussions and for reading and commenting on the draft of our article. It is also a great pleasure to thank G. P. Korchemsky for many useful discussions and comments. We are also grateful to F. Galvagno, M. Preti and K. Zarembo for very interesting discussions.
This research is partially supported by the MUR PRIN contract 2020KR4KN2 ``String Theory as a bridge between Gauge Theories and Quantum Gravity'' and by
the INFN project ST\&FI
``String Theory \& Fundamental Interactions''.

\vskip 1cm

\appendix

\section{Derivation of Eq. \texorpdfstring{\eqref{largeR2p1}}{}}
\label{appendix:A}

In this appendix we argue that at leading order in the strong coupling expansion the function $R_{2p+1}$, defined in \eqref{R2p1}, behaves as follows 
\begin{align}
\label{Rpsc}
R_{2p+1}(\lambda) \underset{\lambda\rightarrow\infty}{\sim} \frac{R_3^{(1)}}{\sqrt{2p+1}}\sum_{\ell=1}^{p}M^{(\infty)}_{2p+1,2\ell+1}\sum_{m=1}^{\ell}\textsf{h}^{(\ell)}_{m}\left(\frac{\sqrt{2m+1}(m^2+m)}{2\,\sqrt{\lambda}}\right) + O\left(\frac{1}{\lambda}\right)\, .
\end{align}
Firstly, let us factorize the part of \eqref{R2p1} which depends on $\lambda$ and define the function
\begin{align}
 F_m^{(p)} \equiv \sum_{n=1}^{\infty}\sqrt{2n+1}\left(\sum_{s=1}^{\infty}\frac{Q_{2s}^{(p)}(n)}{\lambda^{s/2}}\,\textsf{D}_{n,m}\right)\, ,
\end{align}
so that at leading order in the large-$\lambda$ limit it holds
\begin{align}
\label{F11}
F_1^{(1)} = \sqrt{3}\,R_3=\frac{\sqrt{3}\,R_3^{(1)}}{\sqrt{\lambda}}+O\biggl( \frac{1}{\lambda}\biggr)\, .
\end{align}
Now, let us focus on the leading order and make the two following observations. First of all, we notice that the $\lambda$ dependence of the coefficients $\textsf{D}_{n,m}$ is not affected by the value of its indices, namely at leading order these coefficients are always proportional to $\lambda^{-1}$. Indeed as found in \cite{Beccaria:2021hvt}
\begin{align}
\label{eq:D}
& \textsf{D}_{n,m} \underset{\lambda\rightarrow\infty}{\sim} \frac{4\pi^2}{\lambda}\sqrt{(2n+1)(2m+1)} \times   \begin{cases}
n(n+1) \ \ \textrm{for} \ \ n\leq m \\
m(m+1) \ \ \textrm{for} \ \  n \geq m \, .
\end{cases}
\end{align}

Secondly, if we look at the polynomials $Q_{2s}^{(p)}(n)$, we realize that the highest and next-to-highest degree monomials do not depend on $p$ (as one can see in \eqref{Qpoly} for the first polynomials). 

Hence, since we just focus on the leading term of the expansion and we expect that all the remaining lower degree monomials only contribute at subleading orders in the large-$\lambda$ expansion, for any practical purpose we can replace such monomials with the ones with $p=1$. Namely henceforth we identify $Q_{2s}^{(p)} \equiv Q_{2s}^{(1)}$.

These two considerations allow us to conclude that, by construction, at leading order the function $F_m^{(p)}$ must have the same $\lambda$ dependence as the function $F_1^{(1)}$. Therefore we can write
\begin{align}
F_m^{(p)} \underset{\lambda\rightarrow\infty}{\sim} \, f(m)\, F_1^{(1)}\, .
\end{align}
where $f(m)$ does not depend on $\lambda$. Now we determine $f(m)$ which will let us finally prove \eqref{Rpsc}. 

Due to the fact that the leading term in the large-$\lambda$ expansion of $\textsf{D}_{n,m}$ in \eqref{eq:D} behaves differently whether $n\leq m$ or vice versa, it is convenient to rewrite $F_m^{(p)}$ as
\begin{align}
\label{Fmp}
F_m^{(p)}= \sum_{n=1}^{m-1}\sqrt{2n+1}\left(\sum_{s=1}^{\infty}\frac{Q_{2s}^{(p)}(n)}{\lambda^{s/2}}\,\textsf{D}_{n,m}\right)+\sum_{n=m}^{\infty}\sqrt{2n+1}\left(\sum_{s=1}^{\infty}\frac{Q_{2s}^{(p)}(n)}{\lambda^{s/2}}\,\textsf{D}_{n,m}\right)\, .
\end{align}
We observe that, by construction, the second contribution on the r.h.s. has the same large $\lambda$ behaviour as $F_{1}^{(1)}$. On the other hand the finite sum on the r.h.s of \eqref{Fmp} is subleading with respect to $F_1^{(1)}$ because of \eqref{eq:D}.
Therefore we are just left with the series and this quantity must be proportional to $F_1^{(1)}$ through $f(m)$, namely
\begin{align}
\sum_{n=m}^{\infty}\sqrt{2n+1}\left(\sum_{s=1}^{\infty}\frac{Q_{2s}^{(p)}(n)}{\lambda^{s/2}}\,\textsf{D}_{n,m}\right) 
\  \simeq \  f(m)F_1^{(1)} \, .
\end{align}
Using the expressions for the coefficients $\textsf{D}_{n,m}$ \eqref{eq:D} we determine $f(m)$
\begin{align}
f(m) = \frac{\sqrt{2m+1}}{2\,\sqrt{3}}\,m(m+1) \, .
\end{align}
This way we get the final expression for the leading order of the function $F_m^{(p)}$ 
\begin{align}
F_m^{(p)} \underset{\lambda\rightarrow\infty}{\sim} \, \sqrt{2m+1}\,m(m+1)\, \frac{R_3^{(1)}} {2\,\sqrt{\lambda}} + O\left(\frac{1}{\lambda}\right)
\end{align}
and if we substitute this result in \eqref{R2p1} we obtain equation \eqref{Rpsc}.

\printbibliography

\end{document}